\documentclass{aa}

\usepackage{graphicx}
\usepackage{txfonts}

\usepackage[dvipsnames]{xcolor}
\usepackage{xspace}
\usepackage{anyfontsize}
\usepackage{comment}
\usepackage{amsmath}
\usepackage[]{hyperref}
\usepackage[utf8]{inputenc}

\usepackage{upgreek}
\usepackage{natbib}

\newcommand{\decode}{\textsc{decode}\xspace}

\defcitealias{fu_2025}{F25}
\defcitealias{fontanot_2017}{F17}

\begin{document}

   \title{How galaxies acquire their stellar mass at high redshift: High star formation efficiencies and the relative roles of dust and initial mass function}
   \titlerunning{Galaxy stellar mass at high-$z$}
   \authorrunning{H. Fu et al.}

    \author{Hao Fu\inst{1},
            Francesco Shankar\inst{2},
            Fabio Fontanot\inst{3},
            Andrea Lapi\inst{4},
            Feng Yuan\inst{1},
            Mohammadreza Ayromlou\inst{5},
            Daniel Roberts\inst{2},
            Lumen Boco\inst{6},
            Nicola Menci\inst{7},
            Emiliano Merlin\inst{7},
            Laura Pentericci\inst{7},
            Mengyuan Xiao\inst{8}
            }

   \institute{Center for Astronomy and Astrophysics and Department of Physics, Fudan University, Shanghai 200438, China\\
              \email{haofu@fudan.edu.cn}\and
            School of Physics and Astronomy, University of Southampton, Highfield, Southampton, SO17 1BJ, UK\and
            INAF – Astronomical Observatory of Trieste, Via G. B. Tiepolo 11, 34143 Trieste, Italy\and
            SISSA, Via Bonomea 265, 34136 Trieste, Italy\and
            Argelander-Institut f\"ur Astronomie, Auf dem H\"ugel 71, D-53121 Bonn, Germany\and
            Universit{\"a}t Heidelberg, Zentrum f{\"u}r Astronomie, Institut f{\"u}r theoretische Astrophysik, Albert-Ueberle-Str. 2, 69120 Heidelberg, Germany\and
            INAF Osservatorio Astronomico di Roma, Via Frascati 33, 00078 Monteporzio Catone, Rome, Italy\and
            Department of Astronomy, University of Geneva, Versoix, Switzerland
             }

   \date{Received September 15, 1996; accepted March 16, 1997}


%

  \abstract
   {}
   {JWST has measured an unprecedented abundance of galaxies above redshift $z\gtrsim 4-5$, whose formation and evolution are still difficult to reconcile within traditional galaxy evolution models in a Lambda Cold Dark Matter ($\Lambda$CDM) framework. Here we present a study on the star formation histories of these high-redshift galaxies between $z\simeq5-12$ via a cutting-edge data-driven semi-empirical model that uses the observed ultra-violet (UV) luminosity functions (LFs) as input to retrieve star formation rates (SFRs), naturally bypassing any uncertain modelling of cooling, feedback and/or stochastic processes.}
   {Galaxy stellar masses are progressively built in time by integrating their SFRs assigned along their progenitor haloes via the SFR-halo accretion rate relation, derived from abundance matching between the input observed UV LFs with the dark matter halo accretion rate distributions at each redshift. Our original method reverse-engineers empirical estimates of the high-$z$ galactic SFRs directly from observations via abundance matching rather than fitting observed LFs with parametric star formation efficiencies (SFEs). This makes the SFEs a full prediction of the model rather than a tuned input, serving as a solid baseline to test burstiness, dust attenuation, or initial mass function variations.}
   {Our approach reproduces the total stellar mass function, the large-scale clustering, and the star-forming main sequence. We find that massive galaxies grew their stellar mass with a bursty star formation at $z\sim9-10$, broadly in agreement with the star formation histories inferred from spectral energy distribution fitting, with the SFE reaching high peaks of $0.8-0.9$ at $z>9$ and lowering to standard values of $0.2-0.3$ below $z\lesssim9$. We find that the presence of dust could enhance the predicted SFRs at $z\lesssim8$, better reproducing the observed SFRs of massive dusty galaxies, and increase the SFEs to values close to or even above unity at $z \gtrsim 8$. Finally, switching to top-heavy initial mass functions reduces the SFEs by a factor of $2-3$, highlighting the need for a variable initial mass function as an inevitable ingredient in the evolution of galaxies at high redshifts to avoid unphysical SFEs, especially in the presence of dust.}
   {}

   \maketitle
%

\section{Introduction}\label{sec:intro}

The formation and evolution of galaxies are still debated questions today. In a Lambda Cold Dark Matter ($\Lambda$CDM) Universe galaxies grow in stellar mass via both in situ and ex situ processes. The in situ growth is regulated by the star formation rate, i.e. the conversion of cold gas into stars, fuelled by the gas accretion of the host dark matter haloes. The ex situ growth is instead controlled by the rate of mergers with other galaxies, which in a $\Lambda$CDM Universe is the inevitable consequence of the mergers between the host dark matter haloes. Many studies have shown that at redshifts $z \lesssim3$ less massive galaxies (with stellar mass $M_\star \lesssim 10^{11} \, M_\odot$ at $z\sim0$) formed their stellar mass mainly via star formation, whereas more massive galaxies had more contribution from mergers (e.g. \citealt{van_dokkum_2010, shankar_2015, buchan_2016, pillepich_2018, grylls_2019, grylls_2020, fu_2022, fu_2024, eisert_2023}).

At progressively higher redshifts the new observations from the James Webb Space Telescope (JWST) have revolutionised our view of galaxy formation and evolution. Firstly, JWST has revealed a large number of star-forming, ultraviolet (UV)-bright galaxies up to $z\sim17$ and possibly beyond, with abundances exceeding pre-JWST observations by up to an order of magnitude or more (e.g. \citealt{bouwens_2023, finkelstein_2023, finkelstein_2024, adams_2024, mcleod_2024, robertson_2024, perez_gonzalez_2025}). Secondly, a large population of active galactic nuclei (AGN) powered by very massive black holes have also been inferred from JWST observations (e.g. \citealt{matthee_2024, akins_2025, scholtz_2025, juodzbalis_2026}). Last but not least, JWST has also unveiled a population of extremely massive galaxies at $z\gtrsim3$ that were completely missed in previous observations and are challenging to reconcile within current traditional models (e.g. \citealt{donnan_2024, matsumoto_2024, mcgaugh_2024, weibel_2024, shuntov_2025}). For example, \citet{glazebrook_2024} identified a massive quiescent galaxy, ZF-UDS-7329, at redshift $z=3.25$ with inferred stellar mass $M_\star \sim 10^{11.3} \, M_\odot$, speculating that it formed most of its stellar mass at $z\sim 11$ in a burst of star formation. They discuss that this scenario is in tension with current models of galaxy formation within the $\Lambda$CDM framework, pointing possibly towards a different stellar initial mass function (IMF), or changes in the underlying stellar population and feedback modelling. Indeed, evidence for a radially varying IMF in local massive galaxies is mounting (e.g. \citealt{cappellari_2006, conroy_2012, la_barbera_2016, bernardi_2019, marsden_2022, lu_2024}).

More generally, the large number of UV-bright galaxies inferred from JWST observations has been explained in a number of different ways from the theoretical point of view. Some popular explanations rely on stochastic star formation to allow for epochs of larger-than-average galaxy growth rate (e.g. \citealt{shen_2023, sun_2023, gelli_2024}), or reducing the content of dust (e.g. \citealt{ferrara_2024}), lowering stellar metallicity or assuming a top-heavy IMF to boost the UV emission for any given episode of star formation (e.g. \citealt{inayoshi_2022, trinca_2024, yung_2024}), assuming density-modulated star formation efficiencies (e.g. \citealt{somerville_2025}), and/or a feedback-free scenario (e.g. \citealt{dekel_2023}). To explain the formation of massive galaxies at $z>5$ in particular, some groups invoked the extreme value statistics (\citealt{lovell_2023, carnall_2024, jespersen_2025, enriguez_vargas_2026}) to show that the existence of massive galaxies (with stellar mass $M_\star > 10^{11} \, M_\odot$) observed between $3<z<5$ may be accommodated within $\Lambda$CDM but only assuming very high star formation efficiency (SFE; $\epsilon_\star = 0.8-1$). Alternatively, a bottom-heavy IMF could also increase the stellar mass by boosting the fraction of lower mass stars (e.g. \citealt{lapi_2024, yung_2024, mauerhofer_2025, fontanot_2026}). Alternative possibilities for boosting the formation of galaxies at early epochs may be found either in improving the modelling of the halo accretion histories within $\Lambda$CDM (e.g. ellipsoidal collapse) or in varying the underlying cosmological model, such as allowing for dynamical dark energy models (e.g. \citealt{boylan_kolchin_2023, menci_2024, sokoliuk_2025, fakhry_2026, menci_2026}) or modified gravity models (e.g. \citealt{mcgaugh_2024}).

In the context of a $\Lambda$CDM Universe, some groups have attempted to model the galaxy star formation histories and implied UV luminosities via semi-empirical techniques. This approach bypasses the ab initio modelling of the baryonic physics behind star formation in galaxies, which is instead guided by observational data. For example, \citet{kar_2026} and \citet{yung_2025}, along the lines of \citet{moster_2018} and \citet{behroozi_2019}, assumed a parametric SFE of the type $\epsilon_\star (M_{\rm h}, z)$ characterised by a number of free parameters defining its variation as a function of host halo mass and redshift. They concluded that the number density and spatial distribution of UV galaxies can be reproduced within $\Lambda$CDM with relatively standard SFEs of $\sim20-30\%$, but steadily increasing at higher redshifts and without the need for very large scatters of up to $\sim 1$ dex or more in the star formation rate-halo mass relation, as previously inferred by other more comprehensive theoretical models (e,g, \citealt{dekel_2023, mason_2023, nikopoulos_2024}).

In this work we made use, extending it to higher redshifts, of the semi-empirical approach recently devised by our group in two semi-empirical models, \decode (\citealt{fu_2025, fu_2025b}) and \textsc{TopSem} (\citealt{boco_2023}), which have been successful in reproducing the stellar mass functions, star formation histories, and bulge-to-total distributions of galaxies at $z \lesssim 4-5$. Our method relies on extracting the star formation rate-halo accretion rate (SFR-HAR) relation from abundance matching between the SFR and HAR functions, where the former is directly inferred from observations and the latter is theoretically predicted by N-body simulations. There are several key advantages of using this data-driven technique. First, it does not require heavy parametrisations of the SFE, which is instead a natural prediction of the abundance matching procedure. Second, it allows us to extract the star formation histories by tracking the SFRs along the host halo progenitor histories. Third, the SFRs in input are directly inferred from observational data and as such they can be considered an effective balance between any feedback and any ex situ or in situ process (e.g. merger, disc instability, clumpy accretion) that may reduce or enhance the star formation in the galaxy. Finally, the SFRs in input do not require any modelling in terms of stochasticity or feedback, and can in turn be effectively adopted to test the degree of SFE, dust attenuation, or variations in IMF, as we carry out below.

The aim of this paper is twofold. On the one hand, we computed the star formation histories of galaxies along their progenitors to compare with direct observations. On the other hand, we integrated them to infer the implied stellar masses, in particular of massive galaxies, global stellar mass functions, and clustering to pin down the SFEs and impact of including dust attenuation or a variable IMF.

This paper is structured as follows. In Section \ref{sec:method}, we describe the methodology followed in our model. In Section \ref{sec:tng_sfr_har}, we test the SFR-HAR assumption at high redshifts by applying it to the TNG simulation. In Section \ref{sec:results}, we show our results on the galaxy abundances, stellar mass-halo mass connection and star formation histories. We also discuss the effect of choosing a different IMF and applying a dust correction to the SFRs. Finally, in Section \ref{sec:discuss} and \ref{sec:conclusions}, we discuss our results and draw our conclusions. In this work, we adopt the $\Lambda$CDM cosmology with best-fit parameters from \citet{planck2018_cosmo_params} (i.e. $(\Omega_{\rm m}, \Omega_\Lambda, \Omega_{\rm b}, h, n_{\rm S}, \sigma_8) = (0.31, 0.69, 0.049, 0.68, 0.97, 0.81)$) and a \citet{chabrier_2003} stellar IMF.

\section{Methods}\label{sec:method}

In this Section, we describe the implementation of our model \decode\footnote{\href{https://github.com/haofuastro/DECODE2}{https://github.com/haofuastro/DECODE2}} (Discrete statistical sEmi-empiriCal mODEl) and our approach to grow and merge galaxies following the dark matter halo assembly histories. We built up galaxies following their progenitors via input scaling relations between the galaxy SFR and HAR. In the first two papers of the series (\citealt{fu_2025, fu_2025b}) we accurately tested the SFR-HAR abundance matching at redshift $z\lesssim 5-6$. In this paper, we focus on the galaxy star formation rates at high redshifts and how they compare to the JWST spectral energy distribution (SED)-inferred star formation histories at redshift $z \gtrsim 5$.

    \subsection{Dark matter halo merger tree}\label{sec:DM}

    We based our analysis on a catalogue of $n= 5\times10^5$ central dark matter haloes with mass above $M_{\rm h} > 10^{10.5} \, M_\odot$ at each redshift of interest. The choice of the volume and cut in halo mass represents a good balance among computational efficiency, statistics and mass resolution for the stellar mass range of interest. The cut in halo mass in particular ensures a completeness in stellar mass down to $M_\star \gtrsim 3 \times 10^9 \, M_\odot$. These haloes were randomly extracted from the \citet{tinker_2008} halo mass function at any redshift above the halo mass cut-off. We ignored the surviving satellite subhaloes whose abundance at these high redshifts is less relevant. The mass accretion history of each dark matter halo and their merger trees were computed via the SatGen analytical code from \citet{jiang_2021}, based on the \citet{parkinson_2008} algorithm fitted to the outputs of the MultiDark Planck N-body simulation. Each of these parent dark matter haloes comes with a mass accretion history along with the infall redshift of the satellite subhaloes and their accretion history before infall. We also checked that these halo accretion tracks are on average well consistent with those from high-redshift dark matter-only simulations, such as \textsc{gureft} (\citealt{yung_2024_gureft}). On top of this dark matter framework, we self-consistently grew central galaxies by assigning to them a SFR following the growth of the host haloes.

    \begin{figure*}
        \centering
        \includegraphics[width=\textwidth]{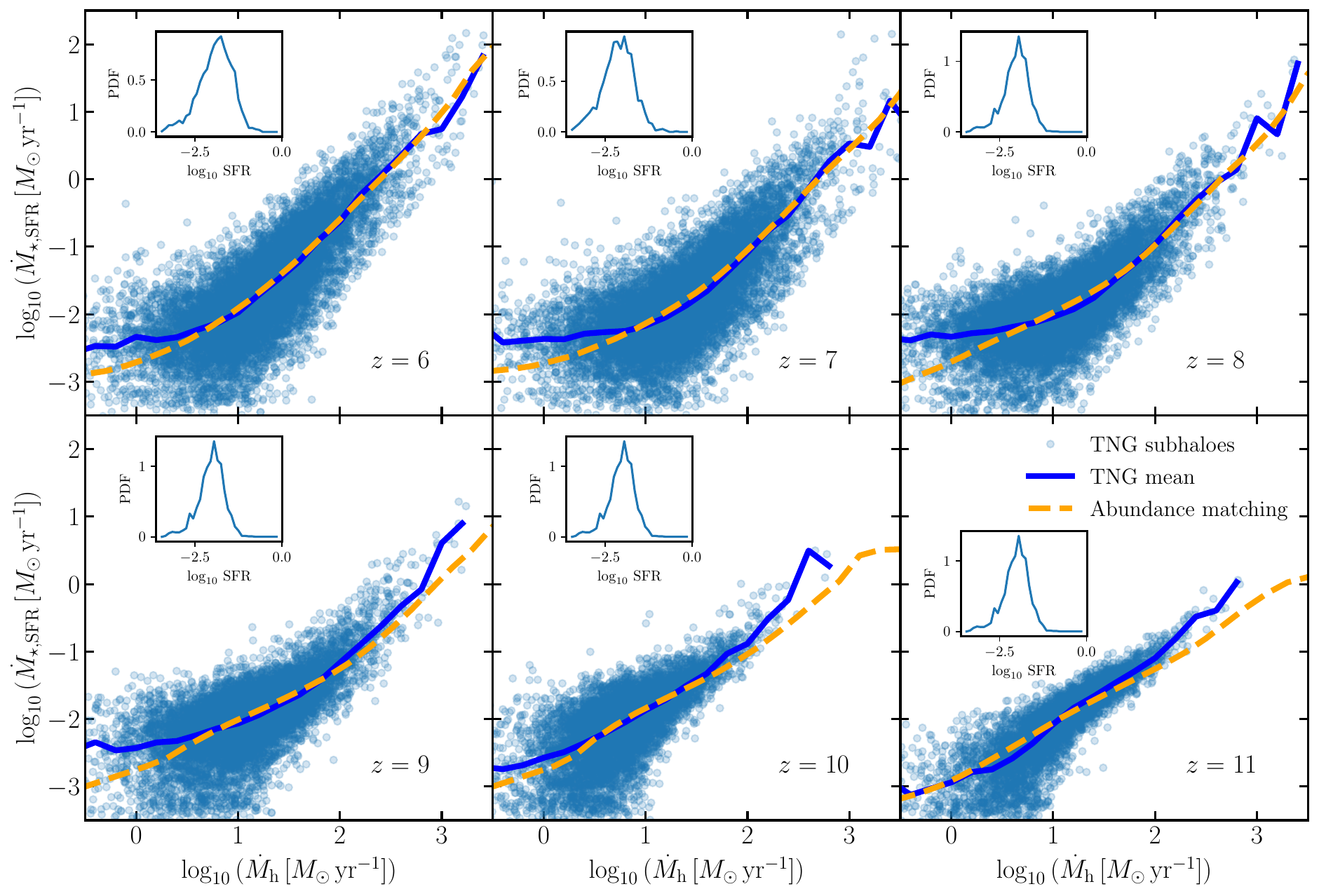}
        \caption{Distribution of the TNG subhaloes on the star formation rate-halo accretion rate plane (blue dots), and median star formation rate-halo accretion rate relations from the TNG (blue lines) and computed using the TNG's inputs via the abundance matching described in Section \ref{sec:AM} (orange dashed lines) at redshifts $z=6-11$. The subplots in each panel show the distribution normalised to unity of the star formation rate in a halo accretion rate bin of $0.2$ dex width centred at $\dot{M}_{\rm h} \sim 10 \, M_\odot \, {\rm yr}^{-1}$, and the distributions in other bins follow a similar behaviour.}
        \label{fg:sfr_har_tng}
    \end{figure*}

    \subsection{Abundance matching}\label{sec:AM}

    The core of our model is based on the abundance matching between the galaxy SFR and host dark matter HAR. To compute the SFR-HAR relation, we performed the abundance matching between the observed SFR function and the theoretical HAR function (we used Equation 37 in \citealt{aversa_2015}, see also \citealt{fu_2022}, \citealt{boco_2023}, \citealt{fu_2024} and \citealt{fu_2025, fu_2025b}). In the abundance matching we assumed a scatter of $0.4$ dex in SFR at fixed HAR, as suggested by our analysis from the TNG simulation. We checked that by varying the value of the scatter within a reasonable range ($0.3-0.5$ dex) our results do not alter appreciably, as also discussed in \citet{fu_2025} and \citet{fu_2025b}, with stellar mass assemblies altering by less than $\lesssim 0.05$ dex.

    \begin{figure*}
        \centering
        \includegraphics[width=\textwidth]{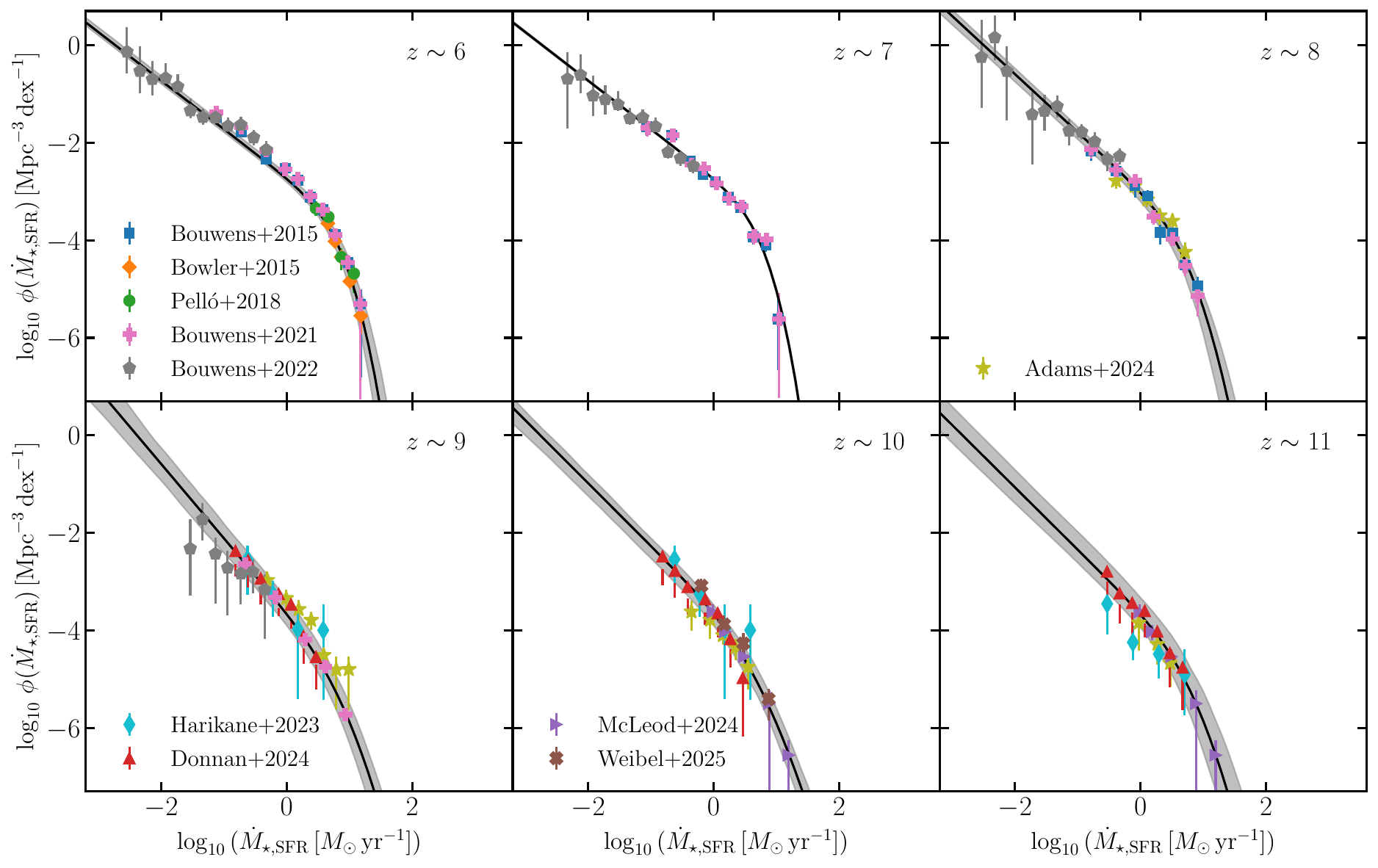}
        \caption{Star formation rate function at from redshifts $z = 6$ to $11$. The data points with error bars show the data from HST (blue squares, pink pluses, and grey pentagons; \citealt{bouwens_2015, bouwens_2021, bouwens_2022}), UltraVISTA/COSMOS and UKIDSS (orange rhombuses; \citealt{bowler_2015}), WUDS (green dots; \citealt{pello_2018}), PEARLS+JWST (yellow stars; \citealt{adams_2024}), JWST-PRIMER (red triangles; \citealt{donnan_2024}), JWST ERO+ERS (cyan rhombuses and purple triangles; \citealt{harikane_2023, mcleod_2024}) and PANORAMIC (brown crosses; \citealt{weibel_2026}). The black solid lines and shaded areas show our fit to Equation (\ref{eq:saunders}) and $1\sigma$ uncertainty.}
        \label{fg:phi_sfr_data}
    \end{figure*}

    In our abundance matching, we input the theoretical HAR function (from the results of the SatGen algorithm) and the observed SFR function (gathered from multiple datasets). For the former, we sampled the distributions from our mock halo catalogue at each redshift, while for the latter, we combined the most recently observed UV galaxy luminosity function (LF) at high redshifts (see Appendix \ref{app:highz_SFRF} for details). We converted UV luminosities to SFRs via the calibration factor from \citet{kennicutt_1998} which yields the following formula (e.g. \citealt{madau_dickinson_2014})
    \begin{equation}
        {\rm SFR}_{\rm UV} =  k_{\rm UV} \cdot L_{\rm UV} \, ,
    \end{equation}
    with $k_{\rm UV} = 2.5 \times 10^{-10} \, M_\odot \, {\rm yr}^{-1} \, L_\odot^{-1}$. We note that the \citet{madau_dickinson_2014} conversion factor is based on a fixed star formation history and solar metallicity, which might systematically overestimate the SFRs up to $1$ dex (e.g. \citealt{wilkins_2020}). However, works on the mass-metallicity relation suggest a metallicity in the range of $Z \in [0.1\,Z_\odot, \, Z_\odot]$ for galaxies with stellar mass $M_\star \gtrsim 10^9 \, M_\odot$ from both JWST emission-line analyses (e.g. \citealt{curti_2023, nakajima_2023, trump_2023, merlin_2025}) and cosmological simulations (e.g. \citealt{wilkins_2023}). We checked that in the most extreme case of lowest metallicity (i.e. $Z=0.1\,Z_\odot$), the UV-SFR calibration is expected to vary by up to $\sim 20 \%$, as suggested by stellar population synthesis models (e.g. \citealt{bicker_2005, wilkins_2019}). Accordingly, the resulting shift in SFR would be less than $<0.1$ dex, corresponding to a $\sim 0.1$ dex shift in stellar mass. For the most massive galaxies, with $M_\star \gtrsim 10^{11} \, M_\odot$, both observations and models suggest a near-solar metallicity (e.g. \citealt{merlin_2025}), implying a negligible alteration in their stellar mass growth.

    Hereafter, we adopt the UV SFR as representative of the total SFR, along with a \citet{chabrier_2003} IMF, as a reference model. We also explore the consequences of considering the dust correction in observed luminosities and adopting different IMFs in Sections \ref{sec:res_dust_corr} and \ref{sec:res_imf}.

    \subsection{Growing galaxies within DECODE}

    We grow our galaxies by assigning to each galaxy a SFR at each redshift via the SFR-HAR relation, following the mass accretion history of the host dark matter haloes. Starting our simulation at redshift $z\sim12$, we then integrate the SFRs across cosmic time
    \begin{equation}
        M_\star (t) = \int_{t_0}^t \dot{M}_{\rm SFR} (t') (1 - \mathcal R) \mathrm dt' \; ,
    \end{equation}
    where $\mathcal R$ is the stellar mass loss fraction that returns into the interstellar medium. We note that the recycling factor $\mathcal R$ is dependent on the IMF which produces variations between $\sim 0.3-0.6$ (see, e.g. \citealt{segers_2016, yu_2016, hopkins_2018}). In what follows, we adopt $\mathcal R = 0.45$, which is a typical value for a Chabrier IMF (e.g. \citealt{reimers_1975, renzini_1988, girardi_2000, pietrinferni_2004, romano_2010, nomoto_2013, vincenzo_2016}). We also clarify that we do not push our simulation beyond $z\gtrsim12$ because the UV LF is not observationally well constrained at those redshifts, even though there are works that have provided preliminary estimates up to $z\sim25$ (e.g. \citealt{donnan_2024, harikane_2025, perez_gonzalez_2025, weibel_2026}).

    \begin{figure}[h!]
        \centering
        \includegraphics[width=0.93\columnwidth]{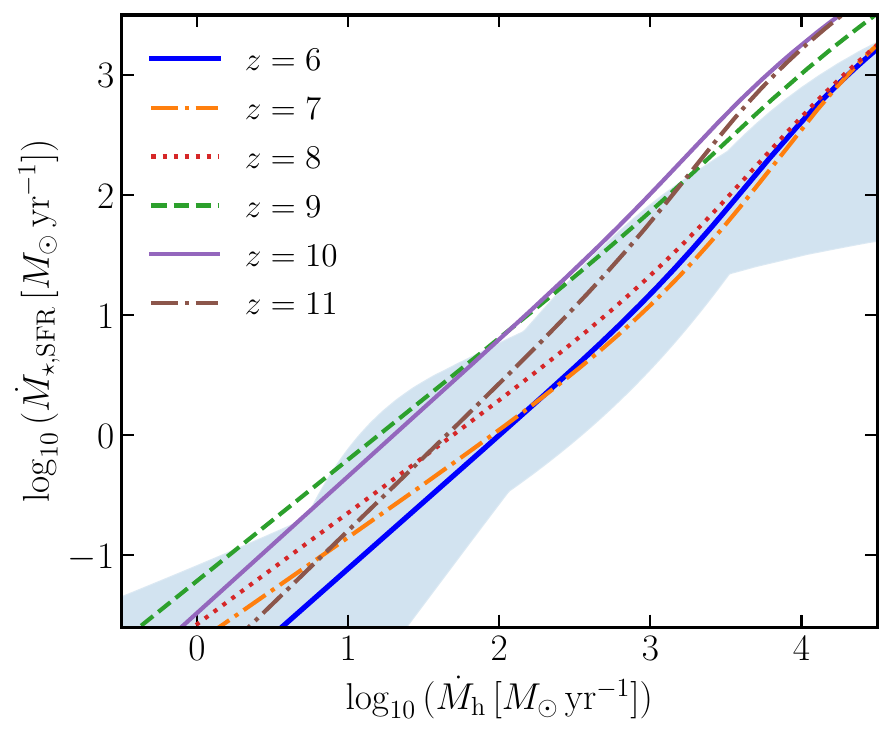}
        \includegraphics[width=0.93\columnwidth]{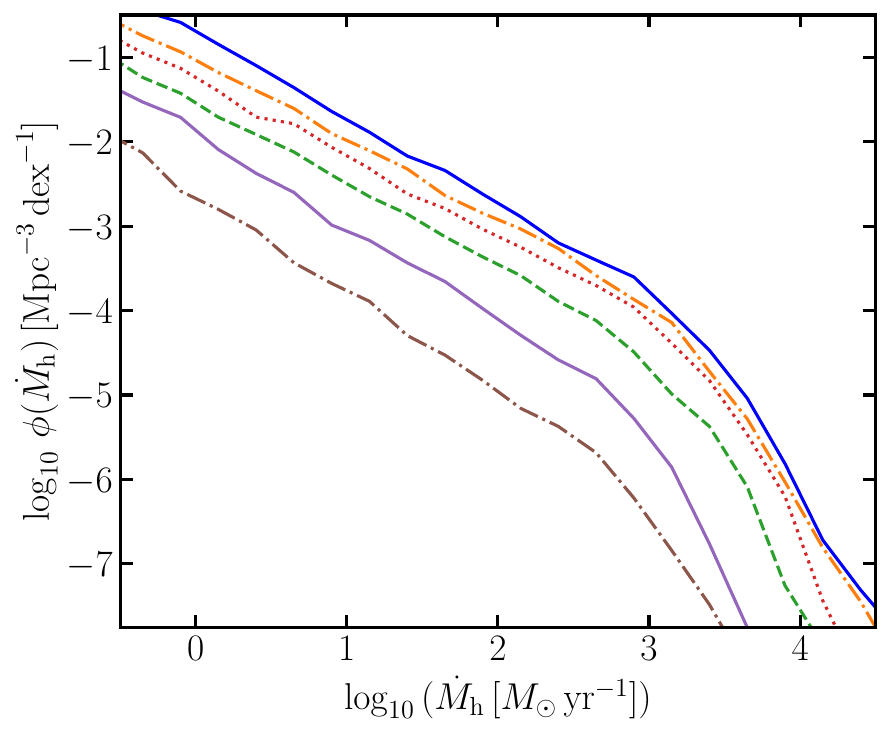}
        \includegraphics[width=0.93\columnwidth]{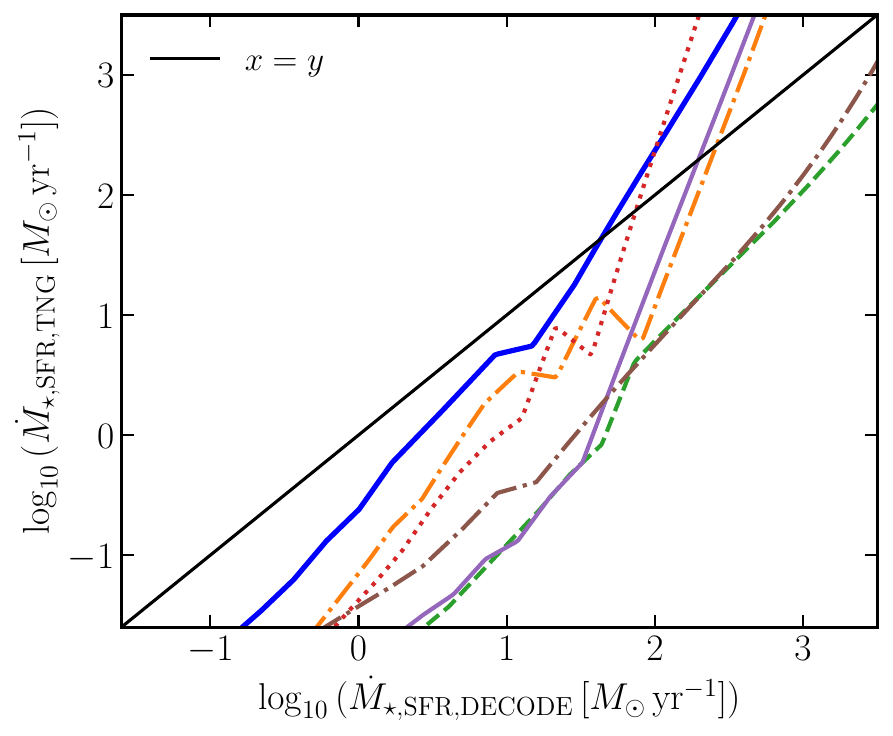}
        \caption{Upper panel: Star formation rate-halo accretion rate relation from redshift $z=6$ to $11$, from the abundance matching. The cyan coloured area shows the redshift evolution region of the relation from $z=0$ to $6$, as presented in \citet{fu_2025} and \citet{fu_2025b}. Central panel: Halo accretion rate function in the same redshift bins. Lower panel: Star formation rate from the TNG simulation compared to that predicted by our abundance matching at the same halo accretion rates.}
        \label{fg:sfr_har}
    \end{figure}

    \subsection{Galaxy mergers}\label{sec:method_mergers}

    We add the contribution from mergers to the total galaxy stellar mass by following the merger histories of the host dark matter subhaloes. In particular, we assign satellite galaxies with a stellar mass at infall via the typical stellar mass corresponding to the halo mass of the host subhalo on the stellar mass-halo mass ($M_\star - M_{\rm h}$) relation of the central counterpart, assuming that satellite galaxies align with the scaling relations of central galaxies at the time of infall. We assume that galaxies merge with a time delay with respect to the host subhaloes following the merger timescales described in Section 3.4 of \citet{fu_2022} corrected by a factor of $\sim 1.5$ at redshift $z\gtrsim5$. These timescales are fitted to reproduce the number densities of surviving satellite subhaloes of the Millennium simulation up to redshift $z\sim 11-12$.

    For the purposes of this paper, we do not discuss in detail the contribution from mergers, even though we include them in the overall stellar mass assemblies of our galaxies for completeness. We have checked that mergers contribute to less than $\sim 5-10\%$ of the total stellar mass growth in all our mock galaxies at $z\gtrsim 5-6$, resulting in a minimal impact on the stellar mass function. We note that this estimate is nicely consistent with the recent observational work by \citet{calabro_2026}, who claimed that mergers can contribute up to $5-10\%$ at $5<z<14$ in a spectroscopic sample of 1233 JWST galaxies. We will dedicate a separate work on the prediction of merger rates of galaxies at high redshifts (Fu et al. in prep.).

    \begin{figure}
        \centering
        \includegraphics[width=\columnwidth]{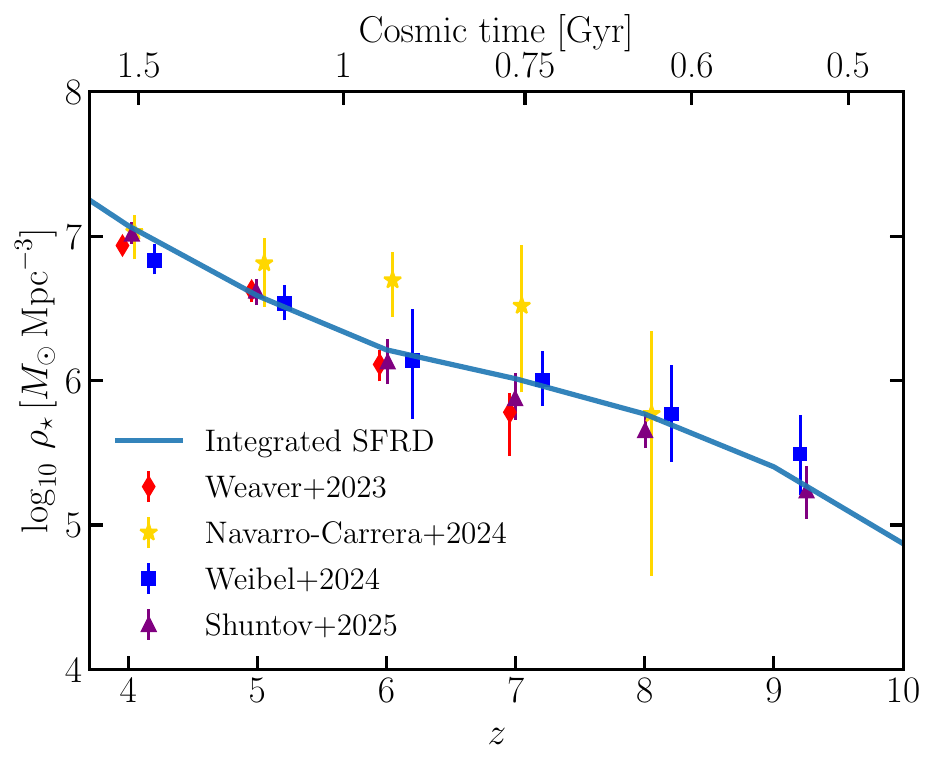}
        \caption{Evolution of the cosmic stellar mass density. The blue solid line shows the cosmic star formation history computed by integrating the cosmic star formation rate density retrieved from our star formation rate functions shown in Figure \ref{fg:phi_sfr_data}. The data points show the observationally inferred stellar mass densities from COSMOS2020 (red rhombuses with error bars; \citealt{weaver_2023}), JWST HUDF and UDS fields (yellow stars with error bars; \citealt{navarro_carrera_2024}), JWST NIRCam (blue squares with error bars; \citealt{weibel_2024}) and JWST COSMOS-Web (purple triangles with error bars; \citealt{shuntov_2025}).}
        \label{fg:cosmic_sfh}
    \end{figure}

\section{Verifying the assumption of a monotonic average relation between SFR and HAR at high redshift}\label{sec:tng_sfr_har}

Before applying our model to predict the star formation histories of high-redshift massive galaxies, following \citetalias{fu_2025} we tested whether the assumption of a monotonic relation between the SFR and the HAR is still expected in cosmological galaxy evolution models even at $z\gtrsim6$. For this purpose, we made use of The Next Generation (TNG henceforth) simulation to test whether a state-of-the-art cosmological hydrodynamic simulation is predicting a monotonic relation between SFR and HAR at $z>6$ as it does at lower redshifts. We stress that this analysis is not aimed at calibrating our model on the TNG simulation which employs its own recipes for the baryonic physics and is calibrated on low-redshift data itself, but only to test the reliability of the SFR-HAR correlation hypothesis in a self-consistent model like the TNG. We note that a whole body of theoretical work supports a close connection between galactic SFR and host HAR (e.g. \citealt{dekel_2009, bournaud_2015, dou_2025, waterval_2025}).

    \subsection{The TNG simulation}\label{sec:tng_description}

    We make use of the cosmological TNG100 simulation, a component of the IllustrisTNG project\footnote{\href{https://www.tng-project.org/}{https://www.tng-project.org/}} (e.g. \citealt{nelson_2019, pillepich_2018a, pillepich_2018, marinacci_2018, springel_2018}). The TNG simulations are performed via the moving-mesh \textsc{arepo} code (\citealt{springel_2010}), employing subgrid physics of galaxy formation such as gas cooling, star formation, stellar evolution, and AGN feedback (e.g. \citealt{pillepich_2018a, weinberger_2017}). The TNG simulations include different volume box sizes and mass relations designed to address specific physical processes. Below, we use the TNG100 simulation run on a 100 Mpc box on a side, which represents a good balance between volume size and mass resolution for the purposes of our test.

    \subsection{The SFR-HAR mapping}\label{sec:sfr_har_tng}

    Figure \ref{fg:sfr_har_tng} shows the SFR-HAR relation above redshift $z\gtrsim 6$, as extracted from the TNG simulation. We find that galaxy SFRs and host HARs are indeed connected via a monotonically increasing relation with a symmetric dispersion around the mean, which can be well approximated by a Gaussian scatter in our abundance matching procedure (as visualised in the inset of each panel in Figure \ref{fg:sfr_har_tng}). We also show that, by taking the TNG's SFR function, HAR function, and scatter as input, our abundance matching (orange dashed lines) is able to reproduce the mean SFR-HAR relation of the simulation out to redshift $z\sim 11$ (blue solid lines), as shown by Figure \ref{fg:sfr_har_tng}. Having confirmed the plausibility of a mean SFR-HAR monotonic relation in a $\Lambda$CDM Universe, along with the validation of our abundance matching procedure to retrieve such a dependence, we are now best positioned to apply this method to the real Universe, with directly observed SFR functions and not derived from a simulation.

\section{Results}\label{sec:results}

In this Section, we present our results on the galaxy stellar mass growths at redshift $z\gtrsim 5$ by adopting as input the SFR-HAR relation from the UV LF abundance matching. In particular, we show the SFR function used as input in our abundance matching and the output SFR-HAR relation (Section \ref{sec:res_sfr_har}). Then we show the predicted stellar mass function (SMF) in Section \ref{sec:res_smf}, and the corresponding $M_\star - M_{\rm h}$ relation and the SFE in Section \ref{sec:res_smhm}. We also show a case study on the star formation histories of the newly detected high-redshift galaxies from JWST (Section \ref{sec:res_sfh}). Finally, we discuss the role of dust obscuration and different IMFs on the galaxy SFRs (Sections \ref{sec:res_dust_corr} and \ref{sec:res_imf}).

    \begin{figure*}
        \centering
        \includegraphics[width=0.32\textwidth]{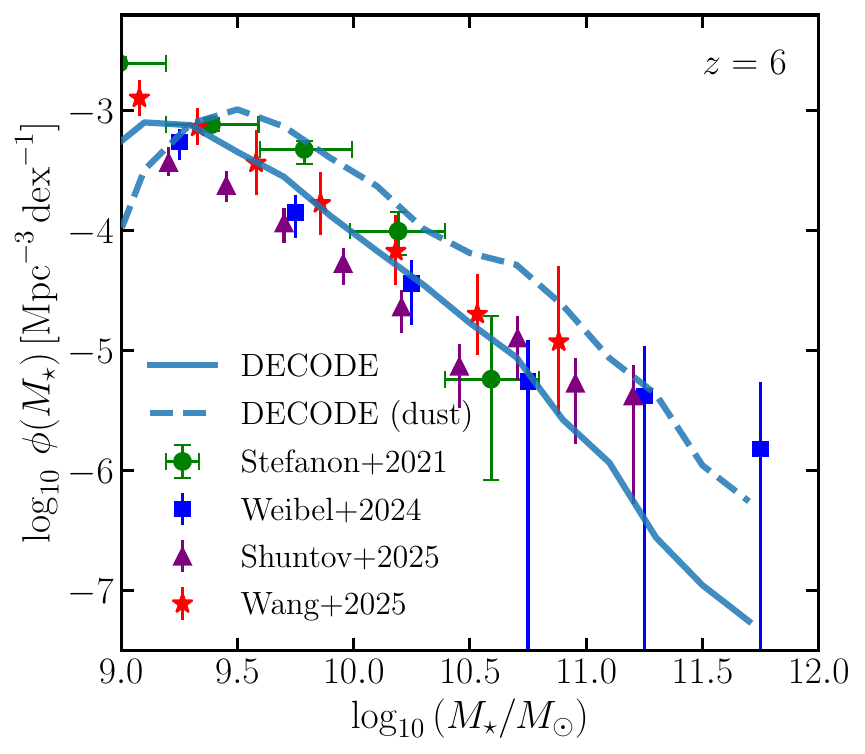}
        \includegraphics[width=0.32\textwidth]{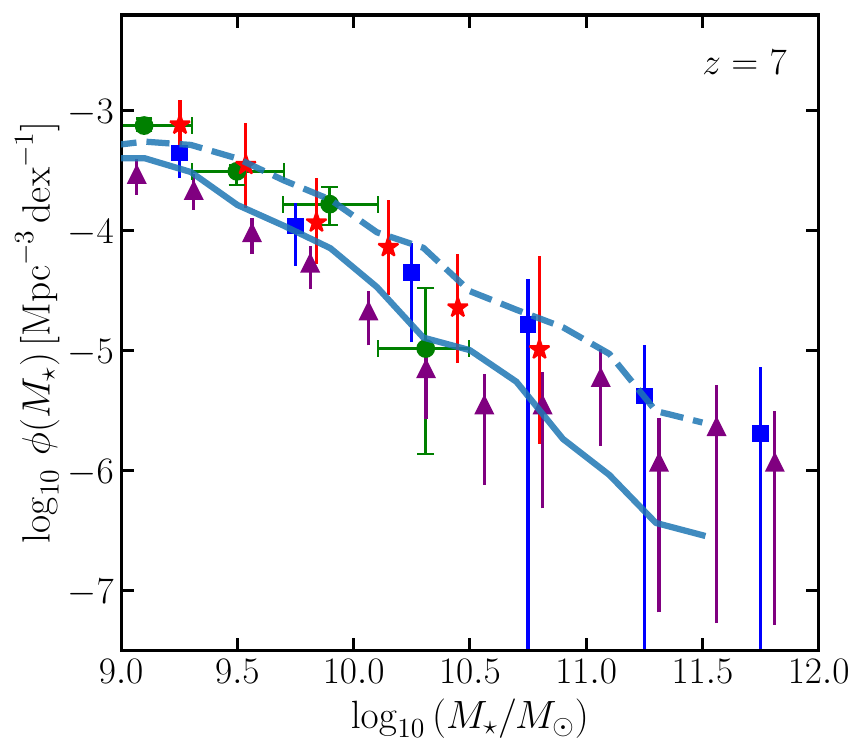}
        \includegraphics[width=0.32\textwidth]{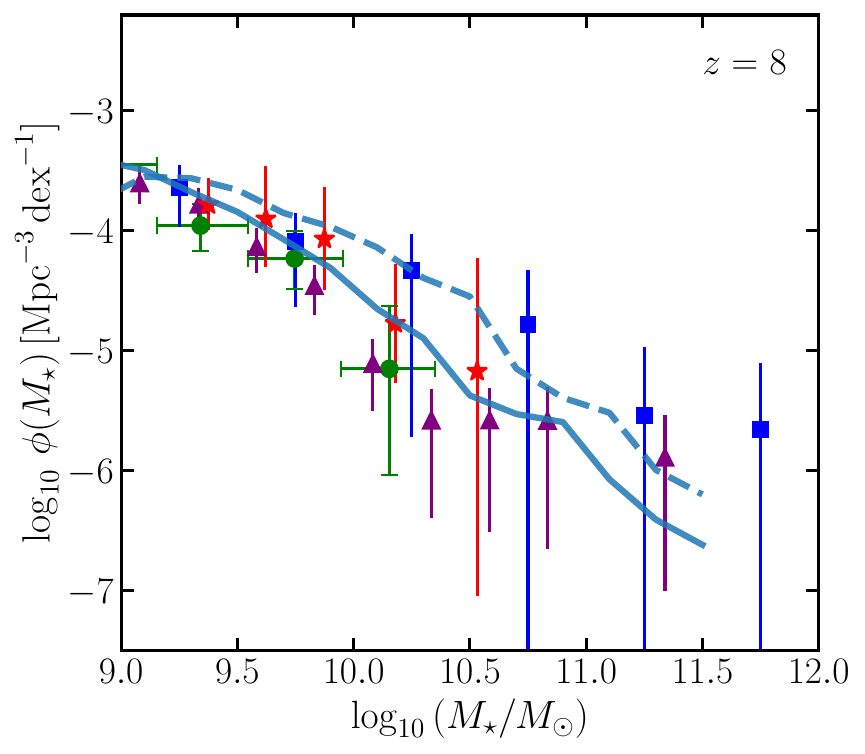}
        \caption{Galaxy stellar mass function at redshifts $z=6$, $7$ and $8$. The blue solid lines show the prediction from \decode's reference model and the blue dashed lines show the scenario with dust correction (as discussed in Section \ref{sec:res_dust_corr}). Our results are compared with the observational data from HST/CANDELS (green dots with error bars; \citealt{stefanon_2021}), JWST NIRCam (blue squares with error bars; \citealt{weibel_2024}), JWST COSMOS-Web (purple triangles with error bars; \citealt{shuntov_2025}), and JWST/MIRI (red stars with error bars; \citealt{wang_2025}).}
        \label{fg:phi_mstar}
    \end{figure*}

    \subsection{Star formation rate-halo accretion rate relation}\label{sec:res_sfr_har}

    We first show the SFR function used in our abundance matching, computed starting from the UV LF, as described in Section \ref{sec:AM}. Within this work we make use of UV data alone since the observational constraints on the infrared (IR) LF are available only up to $z<6$ (e.g. \citealt{gruppioni_2020, wang_2021, traina_2024, traina_2026}). Figure \ref{fg:phi_sfr_data} shows the SFR number densities at redshifts $z=6-11$. In particular, we combine the latest observational determinations in the UV band from HST (\citealt{bouwens_2015, bouwens_2021, bouwens_2022}), UltraVISTA/COSMOS and UKIDSS (\citealt{bowler_2015}), WUDS (\citealt{pello_2018}), PEARLS+JWST (\citealt{adams_2024}), JWST-PRIMER (\citealt{donnan_2024}), JWST ERO+ERS (\citealt{harikane_2023, mcleod_2024}) and PANORAMIC (\citealt{weibel_2026}). We fit the above datasets to the \citet{saunders_1990} analytic form (Equation \ref{eq:saunders}) at each redshift via Markov chain Monte Carlo (MCMC) scans. The best-fitting form to the \citet{saunders_1990} formula is represented by the black lines and shaded areas in Figure \ref{fg:phi_sfr_data}, and the best-fitting parameters with their evolution in redshift are reported in Appendix \ref{app:highz_SFRF}. We find that the observational SFR functions are well described by a power law, characterised by a steep drop towards the bright end ($\dot{M}_{\rm \star, SFR} \gtrsim 3 \, M_\odot / {\rm yr}$). In particular, galaxies with stellar mass $M_\star \sim 10^{11} \, M_\odot$ and SFR in the range $1-3 \, M_\odot / {\rm yr}$ have a number density of $\sim 10^{-5} - 10^{-4} \, {\rm Mpc}^{-3} {\rm dex}^{-1}$ and extremely mild evolution in redshift from $z=6$ to $11$.

    We now turn our attention to the galaxy-halo connection. The upper panel of Figure \ref{fg:sfr_har} shows the SFR-HAR relation at different redshifts, which we stress is a direct result of the abundance matching between our empirical SFR functions and the HAR distributions extracted from merger trees as described in Section \ref{sec:DM}. The resulting SFR-HAR relation at $z\gtrsim 6$ can be well described by a single power law with slope $\sim 1$ and is characterised by a mild time evolution, with increasing normalisation up to $0.5$ dex from $z=6$ to $z=10-11$. This variation in redshift is a direct by-product of the quick HAR function's evolution in normalisation, as shown in the central panel of Figure \ref{fg:sfr_har}, whilst the evolution of the SFR function is weaker across those redshifts. As shown in the lower panel of Figure \ref{fg:sfr_har}, we note that the TNG's SFRs are systematically lower than ours at the same corresponding HARs. This is due to the fact that the TNG physics is calibrated to match the observations at low redshift ($z\lesssim2$) which may not necessarily fit the observed LF at high redshifts, whereas our SFRs come directly from observational constraints. Indeed, at low redshift this difference between simulated and observed SFRs is no longer present. We checked that the origin of this mismatch does not arise from different timescale averages (e.g. \citealt{donnari_2019, donnari_2019_erratum}).

    \subsection{Galaxy stellar mass function}\label{sec:res_smf}

    \begin{figure*}
        \centering
        \includegraphics[width=0.285\textwidth]{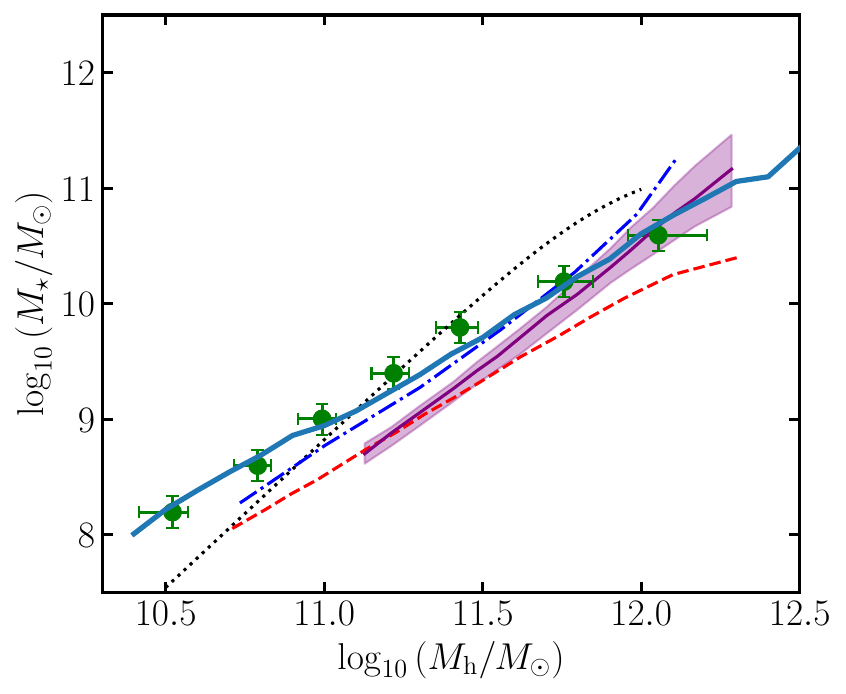}
        \hspace*{0.58cm}
        \includegraphics[width=0.285\textwidth]{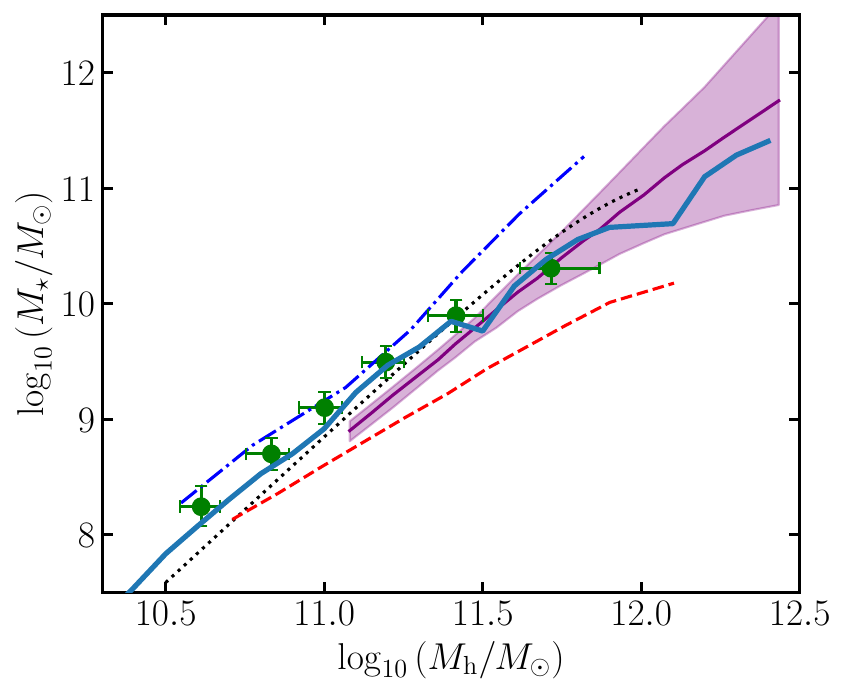}
        \hspace*{0.53cm}
        \includegraphics[width=0.285\textwidth]{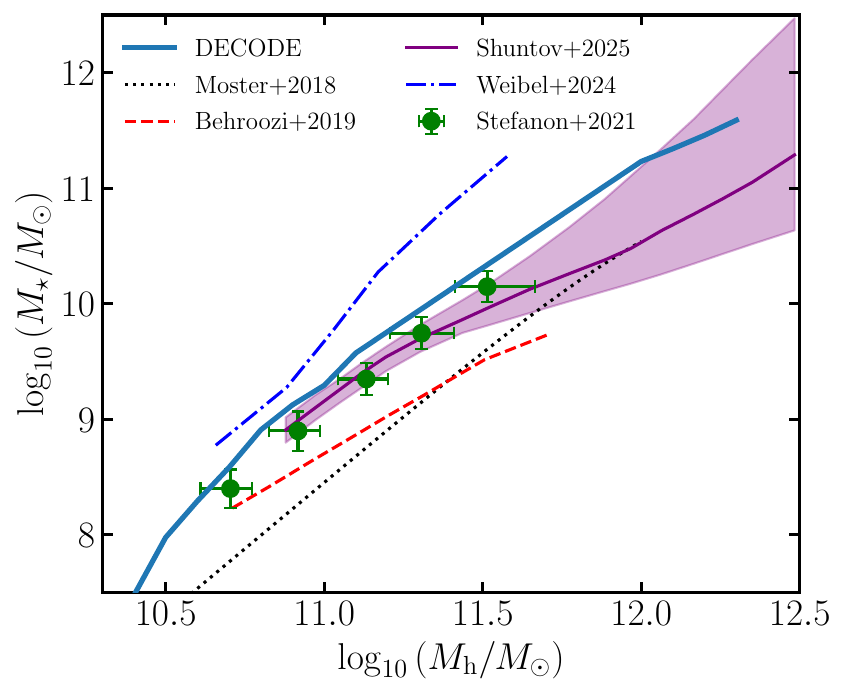}
        \hspace*{0.1cm}
        \includegraphics[width=0.32\textwidth]{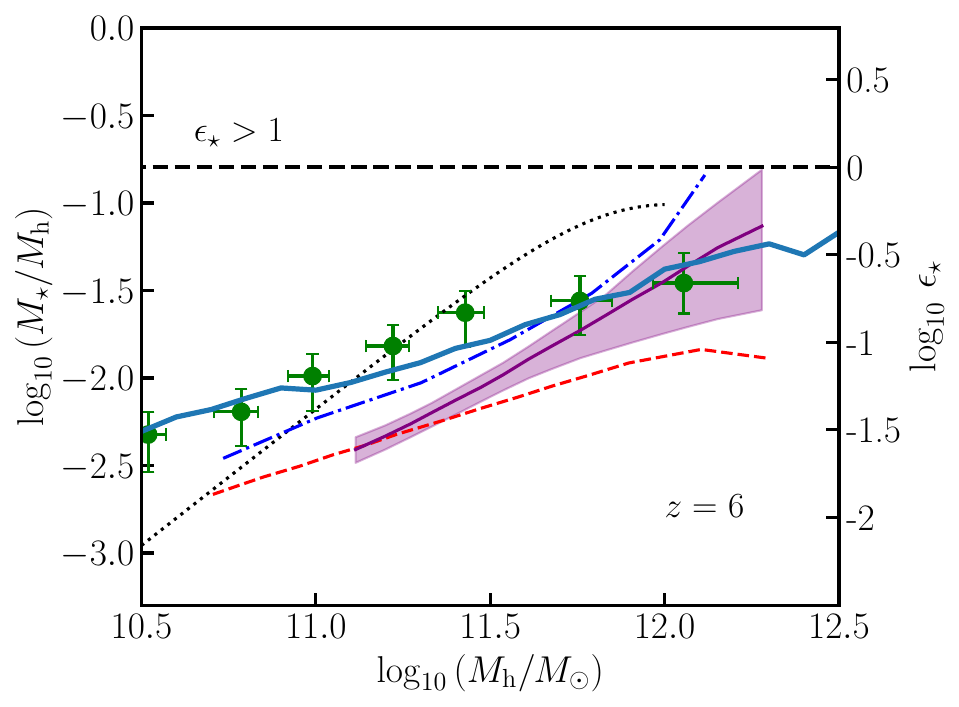}
        \includegraphics[width=0.32\textwidth]{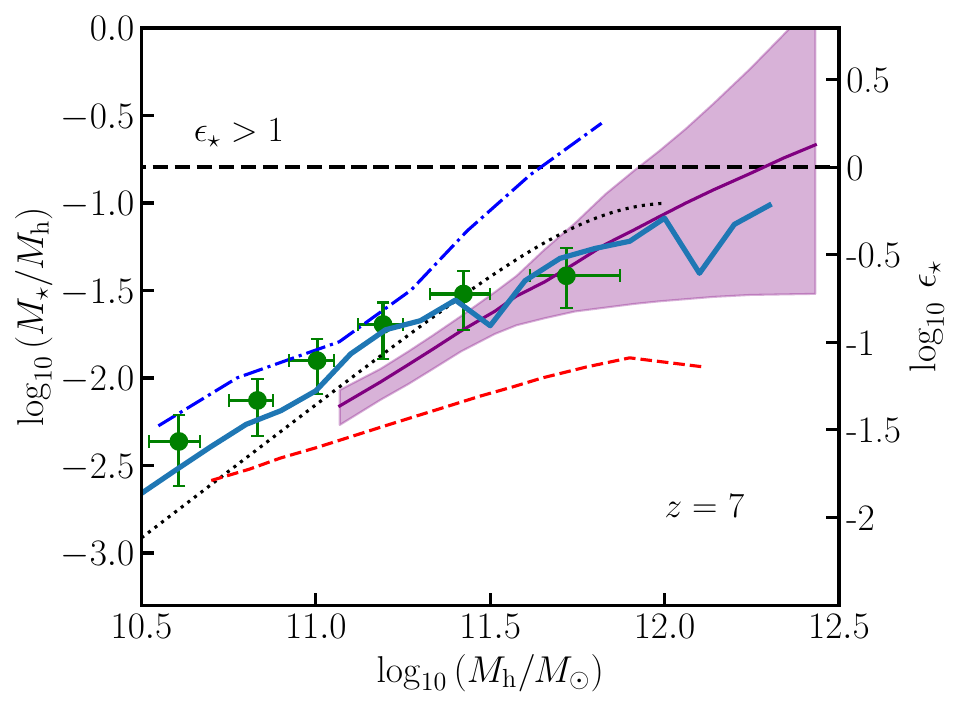}
        \includegraphics[width=0.32\textwidth]{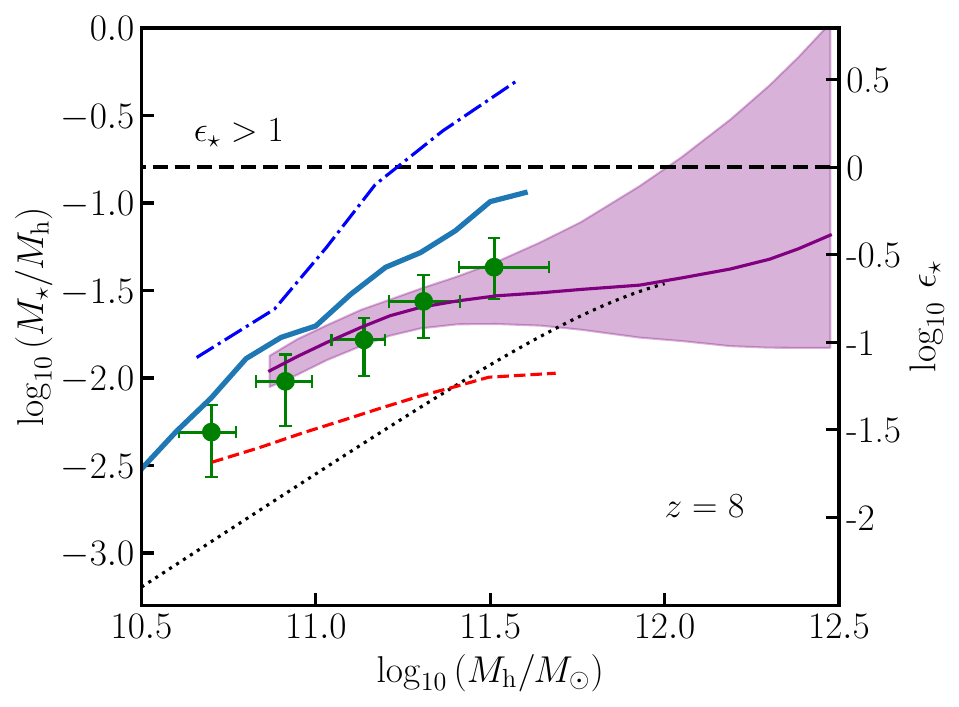}
        \caption{Stellar mass-halo mass relation at redshifts $z=6$, $7$ and $8$. The upper and lower panels show the stellar mass and the stellar-to-halo mass ratio as a function of halo mass, respectively. The right labels in the lower panels represent the star formation efficiency, defined as $\epsilon_\star = M_\star / (M_{\rm h} f_{\rm b} )$. Our results (blue solid lines) are compared to the observational measurements from \citet[][green dots with error bars]{stefanon_2021}, \citet[][purple solid lines with shaded areas]{shuntov_2025}, and to the empirical models of \citet[][black dotted lines]{moster_2018} and \citet[][red dashed lines]{behroozi_2019}. The dark blue dash-dotted lines show the results from the abundance matching using the observed stellar mass function from \citet{weibel_2024}. We note that the stellar mass-halo mass relation might be slightly biased towards higher values below $M_\star \lesssim 10^9 \, M_\odot$, especially at $z\sim 6$, due to the incompleteness of the stellar mass function below that threshold.}
        \label{fg:smhm_sfe}
    \end{figure*}

    In this Section, we compare our SMFs, derived from the time integration of the star formation histories, with those independently measured from COSMOS2020 and JWST at high redshifts. Reproducing the observed SMFs is a crucial self-consistency test to confirm whether our star formation histories are truly representative of the underlying evolution of galaxies. We note that mergers (Section \ref{sec:method_mergers}) and quenching play a minor role in shaping the SMF at $z>6$. Indeed, although our SFR functions are by construction only representative of the star-forming population, the comparison with the total galaxy stellar mass functions is still valid as recent estimates of the quenched fractions at $z\gtrsim5$ is limited to just $ \lesssim 6\%$ for galaxies with for $M_\star \gtrsim 10^{10}\, M_\odot$ (e.g. \citealt{baker_2025, merlin_2025, russell_2025, yang_2026}). At redshifts below $z\lesssim 2-3$ the integral of the observed SFRs produces higher stellar masses with respect to the observed values by a factor of up to $0.5-0.6$ dex, as also found by several works (e.g. \citealt{bernardi_2010, rodriguez_puebla_2017, donnari_2019, leja_2022, fu_2024, bosi_2025}), but we do not show and discuss this any further as it lies beyond the purposes of this work. We also note that satellites contribute $\lesssim 1\%$ to the overall galaxies of $M_\star \gtrsim 10^9 \, M_\odot$ and $z>5$ (Fu et al. in prep.), further validating our assumption of focusing on central galaxies only in this work.

    \begin{figure}
        \centering
        \includegraphics[width=\columnwidth]{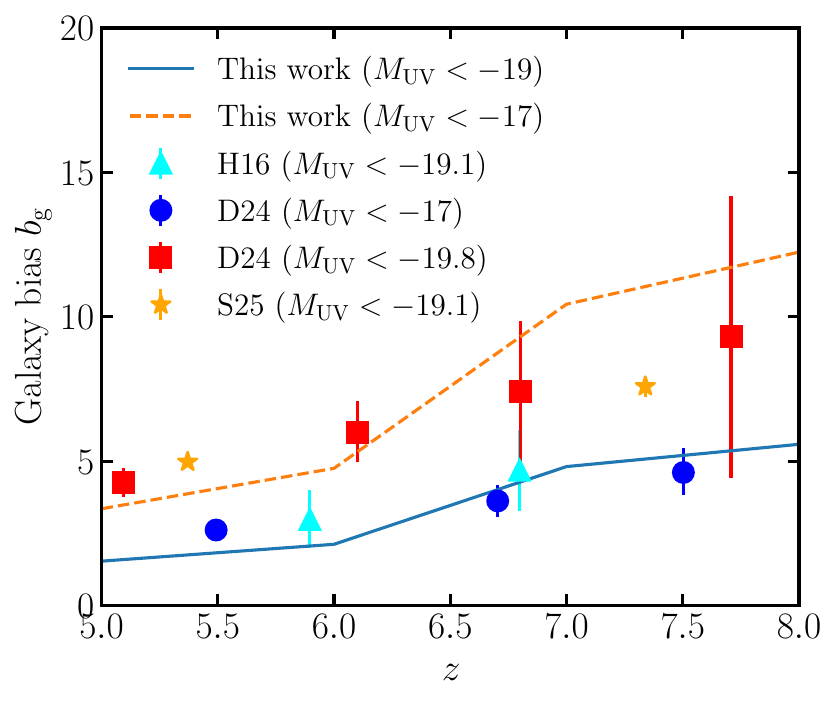}
        \caption{Galaxy bias as a function of redshift for UV magnitude $M_{\rm UV} \lesssim -19$ (blue solid line) and $M_{\rm UV} \lesssim -17$ (orange dashed line). We compare our predictions to the observed galaxy bias from Subaru/Hyper Suprime-Cam (cyan triangles with error bars; \citealt{harikane_2016}), JWST JADES (blue dots with error bars; \citealt{dalmasso_2024b}), HST CANDELS (red squares with error bars; \citealt{dalmasso_2024a}) and JWST NIRCam/grism (orange stars with error bars; \citealt{shuntov_2025b}).}
        \label{fg:bias}
    \end{figure}

    We start by showing in Figure \ref{fg:cosmic_sfh} our integrated stellar mass density (blue solid line) compared against a variety of measurements at $z\gtrsim4$. We find a very good agreement between our predictions and the integrated measurements, thus providing a first validation of our approach. Having compared with integrated quantities, the next step is to generate the full SMF from our model by integrating the SFR of each galaxy along its main progenitor branch. Figure \ref{fg:phi_mstar} shows the galaxy SMF at redshifts $z=6$, $7$, and $8$, as computed from \decode (blue solid lines). We compare our results to the observations from multiple datasets, such as HST/CANDELS (\citealt{stefanon_2021}), JWST NIRCam (\citealt{weibel_2024}), JWST/COSMOS-Web (\citealt{shuntov_2025}), and JWST/MIRI (\citealt{wang_2025}).

    \begin{figure}
        \centering
        \includegraphics[width=0.95\columnwidth]{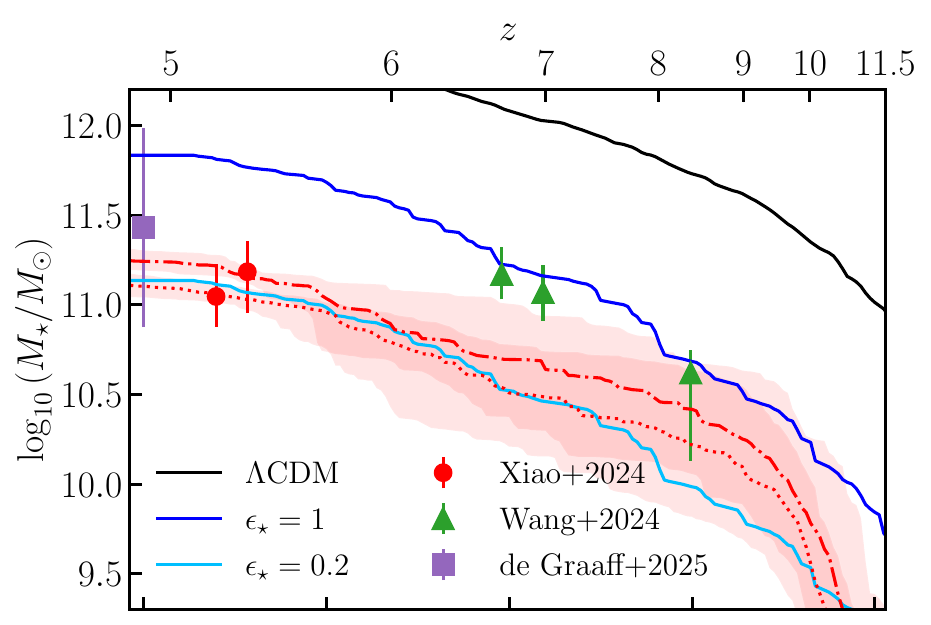}
        \hspace*{0.0001cm}
        \includegraphics[width=0.88\columnwidth]{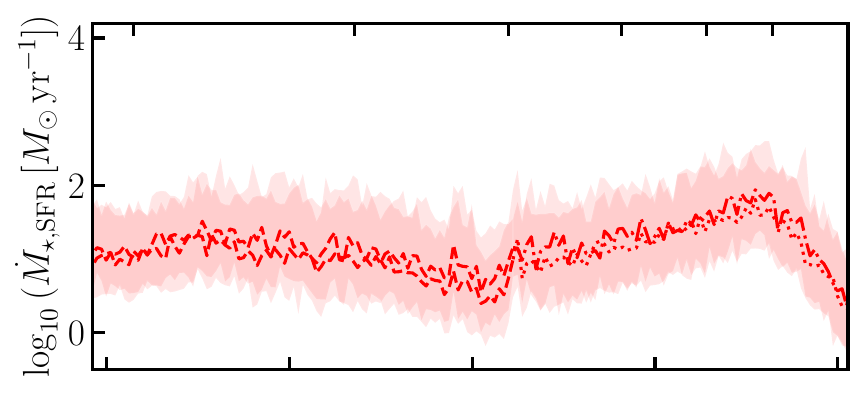}
        \hspace*{-0.4cm}
        \includegraphics[width=0.93\columnwidth]{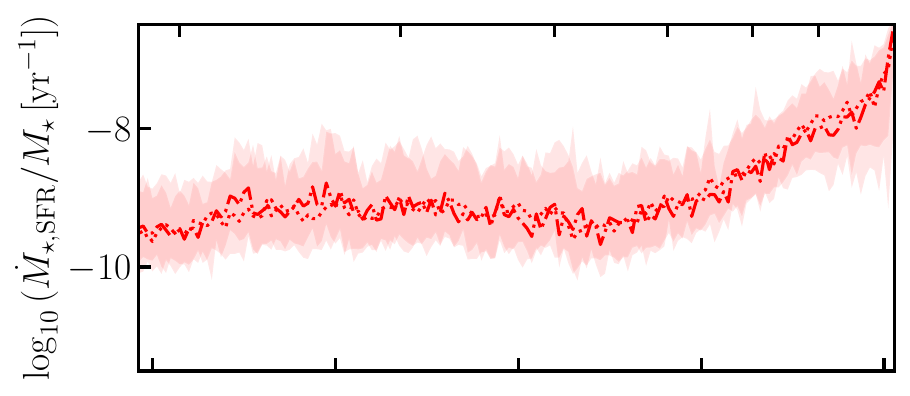}
        \hspace*{-0.25cm}
        \includegraphics[width=0.935\columnwidth]{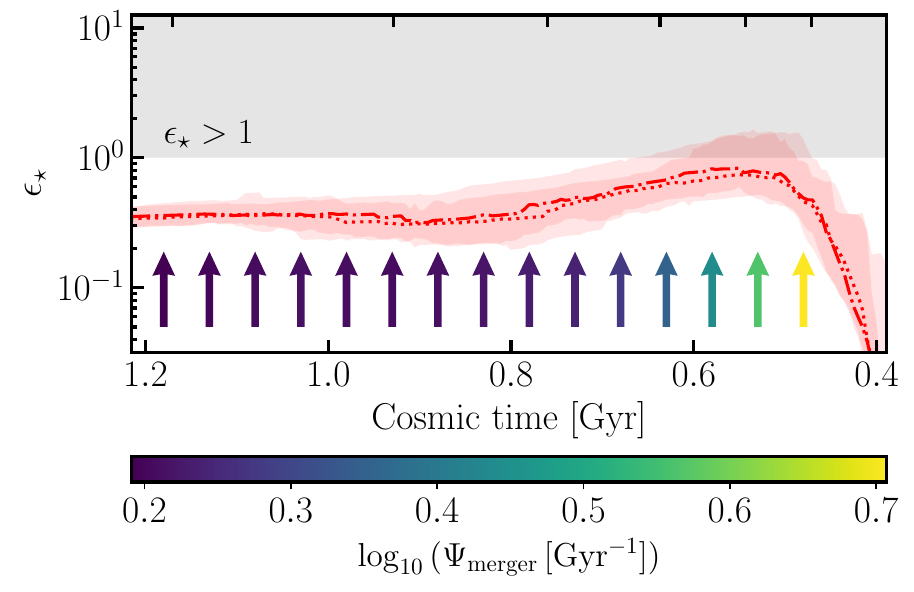}
        \caption{Upper panel: Stellar mass growth of galaxies in different mass bins as a function of redshift. The red dots with error bars show three massive galaxies in the JWST sample from \citet{xiao_2024}. The massive candidates detected by \citet{de_graaff_2025} and \citet{wang_2024_rubies} are also plotted as a reference (purple square and green triangles with error bars). The red lines and shaded areas show the mean stellar mass and $1\sigma$ dispersion in the same mass bins from \decode. The cyan and blue lines show the baryon mass limit computed from the dark matter halo growth with star formation efficiency of $\epsilon_\star=0.2$ and $1$, respectively. The black solid line shows the $\Lambda$CDM limit with $\epsilon_\star = 1 + 3\sigma$. Second, third and fourth panels: Star formation rate, specific star formation rate and star formation efficiency for the same stellar mass bins as in the upper panel. The grey shaded area in the lower panel shows the region where the star formation efficiency exceeds 100\%. The colour-coded arrows in the lower panel represent the merger rate, plotted in terms of number of major mergers per unit time, as a function of redshift.}
        \label{fg:sfh}
    \end{figure}

    Our model can reproduce the bulk of the SMF at all redshifts. This agreement both in normalisation and shape highlights the power of the SFR-HAR connection in tracing galaxy evolution without any further assumption, for example, on the complex and multiple physical processes regulating star formation in galaxies, especially at high redshifts, since the latter are implicitly included in our input SFRs, which should be regarded as a balance among the effects of AGN feedback, stellar feedback, and gas cooling.

    \subsection{Stellar-to-halo connection}\label{sec:res_smhm}

    We now turn our attention to the connection between the galaxy stellar mass and host dark matter halo mass. As discussed by several groups, the shape and evolution of the $M_\star - M_{\rm h}$ relation are directly linked to those of the galaxy SMF (e.g. \citealt{rodriguez_puebla_2017, moster_2018, grylls_2019, fu_2022, fu_2024}).

    \begin{figure}
        \centering
        \includegraphics[width=\columnwidth]{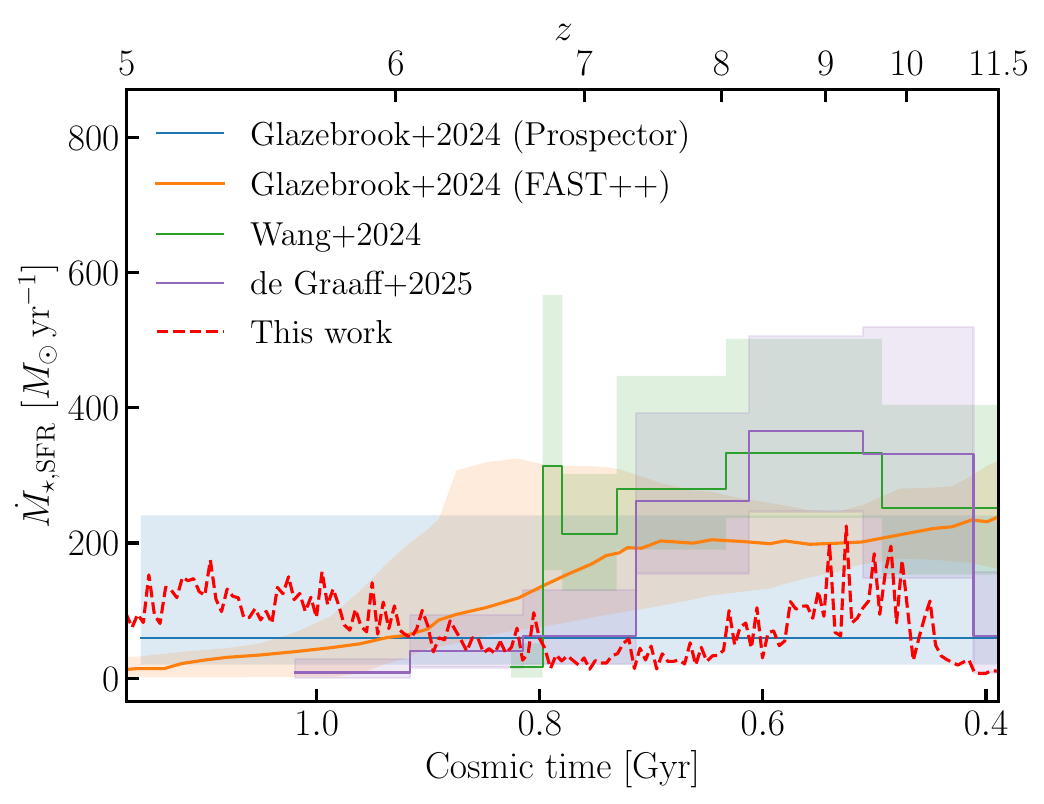}
        \caption{Average star formation rate evolution for galaxies selected in our catalogue with stellar mass $M_\star \sim 10^{11} \, M_\odot$ at redshift $z\simeq5$ (red dashed line). We compare our results to the star formation histories inferred for the JWST ZF-UDS-7329 galaxy from Prospector and FAST++ ($M_\star = 10^{11}\, M_\star$ at $z=3.2$; \citealt{glazebrook_2024}; blue and orange lines with shaded areas), JWST RUBIES-EGS-49140 ($M_\star = 1.5 \times 10^{11}\, M_\star$ at $z=6.7$; \citealt{wang_2024_rubies}; green line with shaded area) and RUBIES-EGS-QG-1 ($M_\star = 10^{11}\, M_\star$ at $z=4.9$; \citealt{de_graaff_2025}; purple line with shaded area) candidates.}
        \label{fg:SFR_evo_data}
    \end{figure}

    Figure \ref{fg:smhm_sfe} displays the $M_\star - M_{\rm h}$ relation at redshifts $z=6$, $7$, and $8$ compared to the measurements from Spitzer/IRAC (\citealt{stefanon_2021}) and COSMOS-Web (\citealt{shuntov_2025}) data. For completeness, we also include the outputs of the empirical models EMERGE (\citealt{moster_2018}) and UniverseMachine (\citealt{behroozi_2019}). The upper panels show the stellar mass as a function of halo mass, and the lower panels report the stellar-to-halo mass ratio and the SFE, defined as $\epsilon_\star = M_\star / (M_{\rm h} f_{\rm b} )$, where $f_{\rm b} = 0.16$ is the cosmic baryon fraction (\citealt{planck2018_cosmo_params}). We find a strong correlation between SFE and halo mass, with a SFE of $\sim70\%$ the baryonic value for the highest masses, significantly above common expectations of $\sim10-20\%$ as calibrated at lower redshifts (e.g. \citealt{hopkins_2011, schaye_2015, pillepich_2018, dave_2019}). We find that our predicted SFEs are well aligned with independent estimates derived from abundance matching of the SMFs and halo mass functions, with a slight tendency for our results to increase by a factor of $\sim2-3$ at $z\geq8$. Our predictions both in stellar mass and in SFE are also higher in normalisation with respect to the results of previous semi-empirical estimates from, e.g.\citet{moster_2018} and \citet{behroozi_2019}, due to the fact that their models were calibrated to match the SMFs and quenched fractions mostly up to $z=4$. Our current predictions are particularly noteworthy as they point to very high baryon conversion efficiencies, particularly in massive haloes at $z\gtrsim7$, in ways independent of any specific theoretical model but simply relying on observed UV data and halo abundances.

        \begin{figure}
            \centering
            \includegraphics[width=\columnwidth]{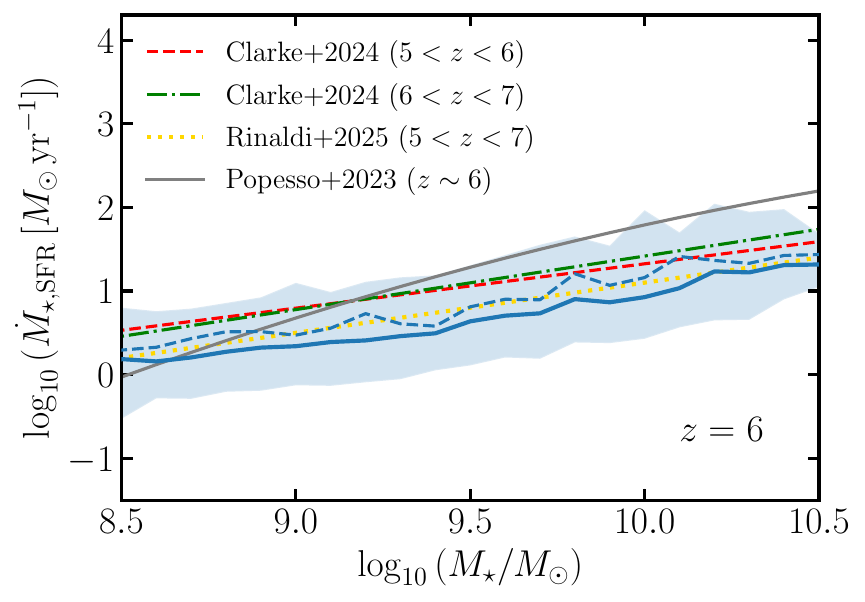}
            \includegraphics[width=\columnwidth]{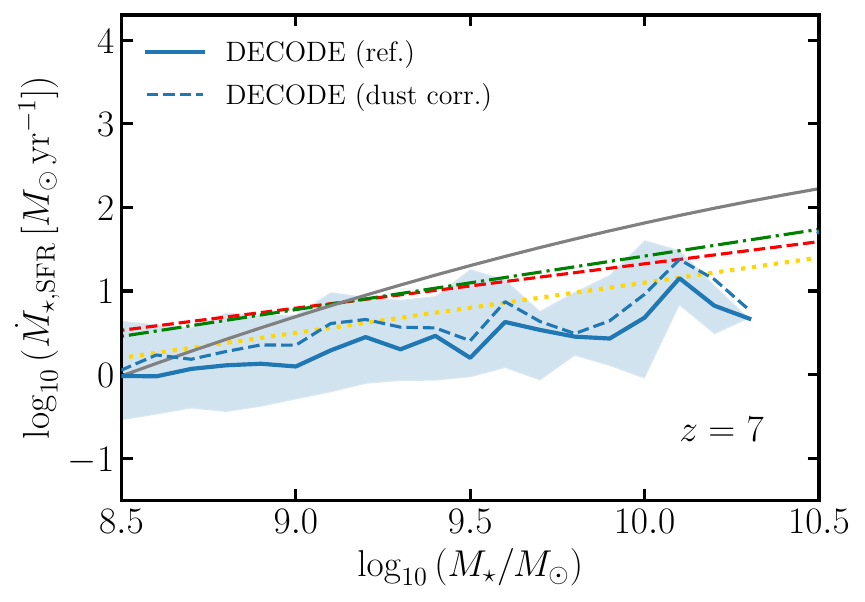}
            \includegraphics[width=\columnwidth]{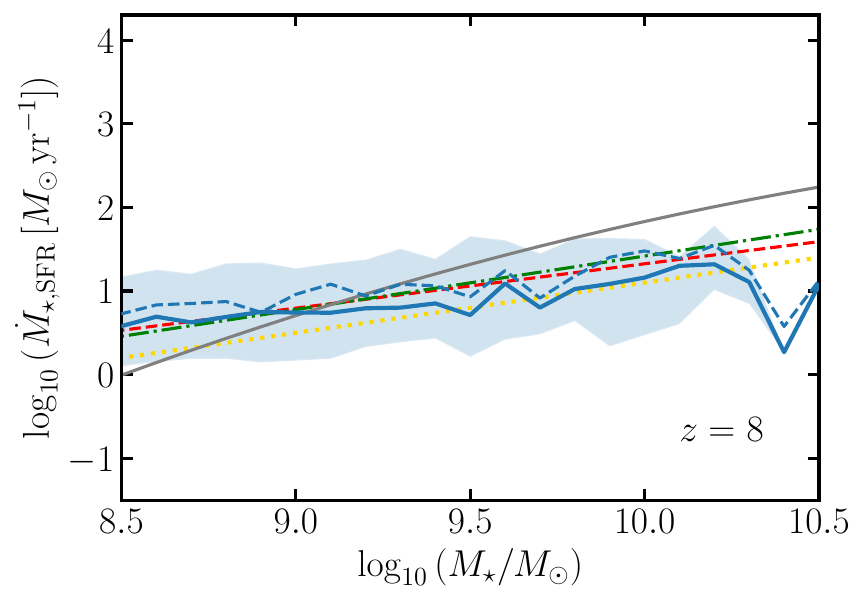}
            \caption{Star formation rate-stellar mass relation at redshifts $z=6$, $7$, and $8$. The predictions from \decode (blue solid lines and shaded areas) are compared to the observational determinations from Spitzer+Herschel (grey solid lines; \citealt{popesso_2023}), JWST JADES/CEERS (red dashed and green dash-dotted lines; \citealt{clarke_2024}), and COSMOS/SMUVS, JWST JADES/GOODS-S and MIDIS/XDF (yellow dotted lines; \citealt{rinaldi_2025}). The blue dashed lines show the model where dust correction is included in the star formation rates.}
            \label{fg:sfms}
        \end{figure}

    \subsection{Clustering}\label{sec:res_clustering}

    Prior to predicting galaxy star formation histories, we computed the large-scale galaxy clustering, as a further test of our approach. Indeed, a robust model of galaxy evolution should be able to reproduce also the clustering in addition to galaxy abundances (e.g. \citealt{mirocha_2020, munoz_2023, gelli_2024, shuntov_2025b, kar_2026}). For this purpose, we calculate the bias as a tracer of the large-scale spatial distribution of galaxies, following the formalism put forward by \citet{munoz_2023}
    \begin{equation}\label{eq:bias}
        b_{\rm g} = \frac{1}{\phi (M_{\rm UV})} \int b(M_{\rm h}) P(M_{\rm h} | M_{\rm UV}) \frac{\mathrm d n}{\mathrm d M_{\rm h}} \mathrm{d} M_{\rm h} \, ,
    \end{equation}
    where $\phi (M_{\rm UV})$ is the UV LF, $b(M_{\rm h})$ is the halo bias (\citealt{tinker_2010}), $P(M_{\rm h} | M_{\rm UV})$ is the probability distribution of halo mass at fixed UV magnitude, and $\mathrm d n / \mathrm d M_{\rm h}$ is the halo mass number density.

    Figure \ref{fg:bias} shows the galaxy bias evolution as predicted by our model, for $M_{\rm UV} \lesssim -19$ and $M_{\rm UV} \lesssim -17$, compared to a variety of observational measurements (\citealt{harikane_2016, dalmasso_2024a, dalmasso_2024b, shuntov_2025b}). Our model predicts an increasing trend for the galaxy bias as a function of redshift and as a function of luminosity at fixed epoch, in good agreement with the data and theoretical expectations of structure formation.

    \subsection{Star formation histories}\label{sec:res_sfh}

    As anticipated above, our methodology allows to track the star formation of galaxies along the progenitors to build full star formation histories and integrated stellar masses. In this Section, we explore the performance of our semi-empirical model in reproducing very massive galaxies of the order of $M_\star > 10^{11}\, M_\odot$ at $z\simeq 5-6$, as observed by \citet{xiao_2024, xiao_2026}.

    Figure \ref{fg:sfh} shows the star formation histories from \decode in different stellar mass bins compared to two massive star-forming candidates detected by JWST (\citealt{xiao_2024, wang_2024_rubies, de_graaff_2025} ). In particular, we show the stellar mass (upper panel), SFR (second panel), specific SFR (third panel), and SFE (lower panel) as a function of redshift. Our model naturally predicts that massive galaxies comparable to those observed by \citet{xiao_2024} underwent a high star formation phase at redshift $z>9$, with a peak of SFR at $z\sim 10$. In particular, these galaxies formed half of their stellar mass in less than 100 Myr with a bursty star formation. The first peak in SFR at redshift $z\sim10$ is characterised by values of SFR between $100-200 \, M_\odot / {\rm yr}$ and SFE close to unity. After $100-150$ Myr of intense star formation the galaxies entered a slightly milder phase with lower values of (specific) SFR and SFE, reaching standard values of $\sim 30\%$ below $z\lesssim7$. As shown in the lower panel of Figure \ref{fg:sfh}, massive galaxies experience more frequent major mergers with increasing redshift (with $\Psi_{\rm merger} \gtrsim 2 / {\rm Gyr}$ at $z\gtrsim8$), suggesting that part of the high SFR before the peak may be ex situ triggered from major mergers. Such a correspondence between high major merger rates and SFRs, expected from the SFR-HAR relation, suggests that major mergers may have contributed to boost early and high SFRs, even if their direct contribution of ex situ stellar mass may have remained relatively minor.

    We now compare our predictions for the SFR evolution to those inferred from SEDs. Figure \ref{fg:SFR_evo_data} compares the mean star formation history from our catalogue to those inferred for massive candidates from JWST (\citealt{glazebrook_2024, wang_2024_rubies, de_graaff_2025}). We selected galaxies of stellar mass $M_\star \sim 10^{11} \, M_\odot$ at $z=5$ to be as consistent as possible with the mass and spectroscopic redshift of the samples from the literature. Our results are broadly in agreement with the star formation histories from SED fitting, yielding a high burst of star formation $\sim 400$ Myr after the Big Bang ($z\sim 11$), with peaks reaching $150-200 \, M_\odot /{\rm yr}$. Interestingly, such SFR quickly drops after $100-150$ Myr, bringing the galaxy to a quiescent mode and after $\sim200$ Myr further resuming to appreciably higher values of the SFR. Such a breathing alternate phases of SFR and quenching have been recently put forward by \citet{merlin_2025} to explain the puzzling non-linear evolution of the relative number densities of quenched and star-forming galaxies at high redshift from ASTRODEEP-JWST. It is interesting to note that we do not include any quenching mechanism in our model galaxies and their apparent periodic behaviour in SFR with the low-SFR phase is uniquely driven by the evolution of the underlying HAR, on the assumption that the host halo (characterised by masses well below the quenching scale of  $M_{\rm h} \sim 10^{12} - 10^{13}\, M_\odot$, e.g. \citealt{dekel_birnboim_2006}) continues carrying cold gas to fuel the star formation in the central galaxy.

    \subsection{Star-forming main sequence}\label{sec:res_sfms}

    In this Section, we turn our attention to the star-forming main sequence of galaxies as predicted by our model. The three panels of Figure \ref{fg:sfms} show the SFR-stellar mass relation at redshifts $z=6$, $7$ and $8$, as labelled, compared to the observed relations from Spitzer+Herschel (\citealt{popesso_2023}), JWST JADES/CEERS (\citealt{clarke_2024}), and COSMOS/SMUVS, JWST JADES/GOODS-S, and MIDIS/XDF (\citealt{rinaldi_2025}). The overall relation is consistent in slope with the \citet{clarke_2024} and \citet{rinaldi_2025} determinations at all redshifts. We also find a mildly increasing normalisation of the SFR as a function of redshift, in agreement with the increase in accretion rate of the host dark matter halo. The good agreement with the main sequence further strengthens the validity and overall self-consistency of our semi-empirical approach.

    \subsection{The role of dust obscuration}\label{sec:res_dust_corr}

    As described above, our SFRs are empirically based on direct observation of the UV luminosities, under the assumption that the dust content at high redshift is low (e.g. \citealt{pettini_1997, nath_2023, shchekinov_2025}). However, the SFRs may still be underestimated due to the effect of dust obscuration that could damp the intrinsic UV luminosity arising from a galaxy (e.g. \citealt{vijayan_2026}), and mounting evidence suggests that the dust may still be present even in very high-redshift galaxies (e.g. \citealt{ferrara_2024, somerville_2025, rodighiero_2026}). Here, we tested the effect of dust obscuration by assuming different attenuation recipes from the literature. In particular, we explored the attenuation laws from \citet{charlot_fall_2000}, \citet{calzetti_2000} and \citet{salim_2018}, which we describe in detail in Appendix \ref{app:dust_laws}. In the same Appendix we also show that the dust-corrected UV LF would match the IR LF at $z \lesssim 6$, where IR measurements are available, providing independent support to our dust-correction recipes.

    We find that within the UV luminosity range explored here, all the recipes quoted above produce similar results, yielding SFRs higher by a factor of $\sim 2-3$ with respect to those predicted by our reference model. This shift would in turn enhance our predicted SFRs for the JWST massive candidates from $\sim 150-200 \, M_\odot / {\rm yr}$ to $\sim 300-600 \, M_\odot / {\rm yr}$, in closer agreement with what measured by \citet{xiao_2024} at redshift $z\sim 5-6$, although still lower than their quoted value of $\sim 700 \, M_\odot / {\rm yr}$. When including dust corrections, the galaxy SMF, shown in Figure \ref{fg:phi_mstar}, is shifted towards higher stellar masses, though still broadly aligned with current data, except possibly at $z\sim6$ and at higher stellar masses. Also the star-forming main sequence, reported in Figure \ref{fg:sfms}, increases by $0.2-0.3$ dex after dust correction. Intriguingly, the dust correction would enhance the SFRs in Figure \ref{fg:sfh} at redshifts $z\sim 8-10$ pushing to SFEs above unity, as we further discuss below in Figure \ref{fg:SFH_SFE_varIMF}. These findings would imply 1) that the dust content may still be low at these high redshifts and high stellar masses and/or 2) that a variable IMF is an inevitable ingredient in the evolution of galaxies at high redshifts. We explore the impact of a variable IMF in the next Section.

    \subsection{The dependence on the initial mass function}\label{sec:res_imf}

    We now discuss the implications of varying the input IMF on the stellar masses and SFRs of the galaxies analysed in this work. The shape of the IMF and its dependence on the galaxy SFR continue to be a matter of intense scrutiny and with the advent of JWST data the topic has become even more relevant. Several works have pointed out the dependence of the stellar population and the derived total stellar mass on the assumed IMF, with the new observed massive candidates pointing towards more exotic, e.g. top-heavy, IMFs (e.g. \citealt{steinhardt_2023, carnall_2024, lapi_2024, van_dokkum_2024, wang_2024a}). Below, we employ two analytical recipes for correcting the galaxy SFR for different assumed IMFs. Our aim is to test whether a variable IMF can relax the pressing need for extreme SFE values in forming massive galaxies. For example, a top-heavy IMF would generate more massive stars increasing the UV luminosity for a given burst of star formation, reducing the need for a high efficiency of star formation.

    First, following \citet{lapi_2024}, we parametrise the IMF via the \citet{larson_1998} analytic formula
    \begin{equation}\label{eq:larson_imf}
        \phi (m_\star) \propto m_\star^\xi \cdot e^{-m_{\rm \star,c} / m_\star} \; ,
    \end{equation}
    where $\xi$ is the slope at high star masses and $m_{\rm \star,c}$ is the characteristic mass below which the IMF flattens or bends. In order to compute the $k_{\rm UV}$ conversion factor from UV luminosity to SFR, we make use of the results from \citet[][see Section 2.2 therein]{lapi_2024} obtained using the \textsc{parsec} code for stellar evolution (\citealt{bressan_2012, goswami_2022}).

    We also take advantage of the predictions of the GAlaxy Evolution and Assembly (\textsc{gaea}) semi-analytical model implementing variable IMF prescriptions. In particular, we consider the \textsc{gaea} realisation discussed in \citet[][F17 hereafter, see also \citealt{fontanot_2024}]{fontanot_2017} including the integrated galaxy-wide IMF (IGIMF) model proposed by \citet{weidner_2005}. The IGIMF framework provides an estimate for the IMF shape for any given reference SFR. In particular, we use the same broken power law parametrisation for the IMF shape adopted in \citetalias{fontanot_2017}, and the corresponding best-fit parameters from their Table 1. The resulting estimates for the $k_{\rm UV}$ factors in the different scenarios allow us to explore a broader, but still physically plausible, range of conversions from UV LF to the SFR function, and probe their impact on our results compared to a Milky Way-like IMF. We show in Appendix \ref{app:SFRF_varIMF} the resulting SFR functions.

    \begin{figure}[h!]
        \centering
        \includegraphics[width=\columnwidth]{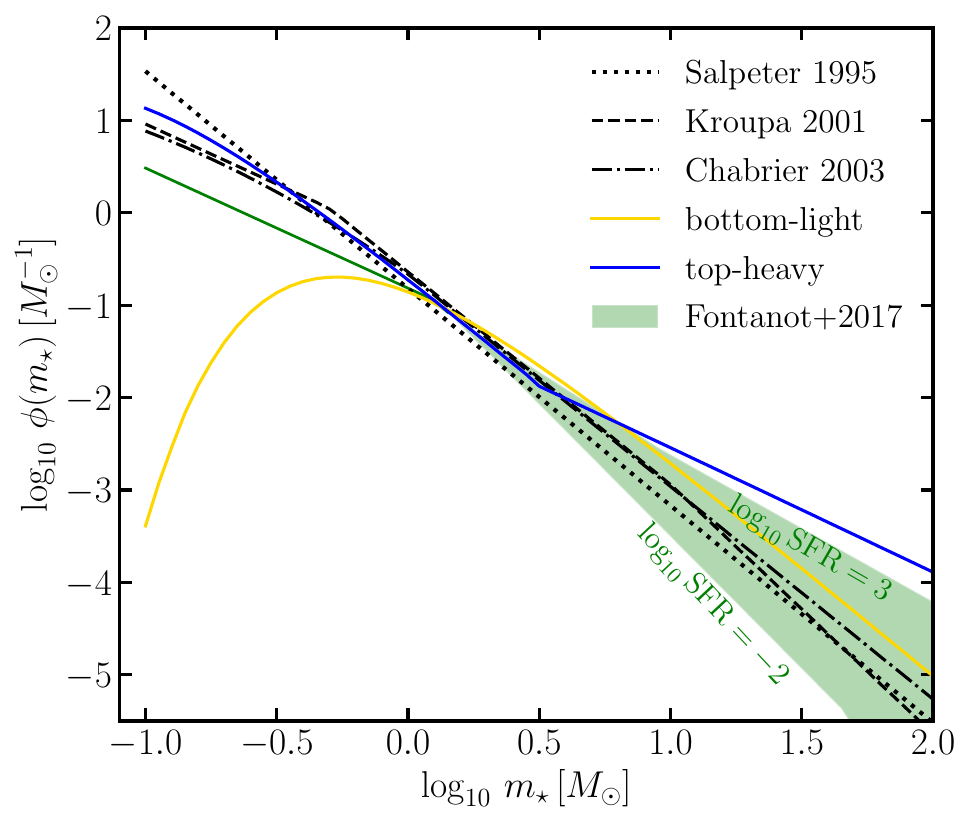}
        \caption{Initial mass function from \citet[][black dotted line]{salpeter_1955}, \citet[][black dashed line]{kroupa_2001} and \citet[][black dash-dotted line]{chabrier_2003}, compared to the bottom-light (gold solid line) and top-heavy (red solid line) toy model IMFs parametrised in terms of the \citet{larson_1998} formula. The green shaded area represents the range of variation of the integrated galaxy-wide initial mass function for star formation rates between $10^{-2}$ and $10^3 \, M_\odot \, {\rm yr}^{-1}$ from \citetalias{fontanot_2017}. The different IMFs are normalised to the value of the \citet{chabrier_2003} IMF at $m_\star = 1 \, M_\odot$.}
        \label{fg:IMFs}
    \end{figure}

    \begin{figure}
        \centering
        \includegraphics[width=\columnwidth]{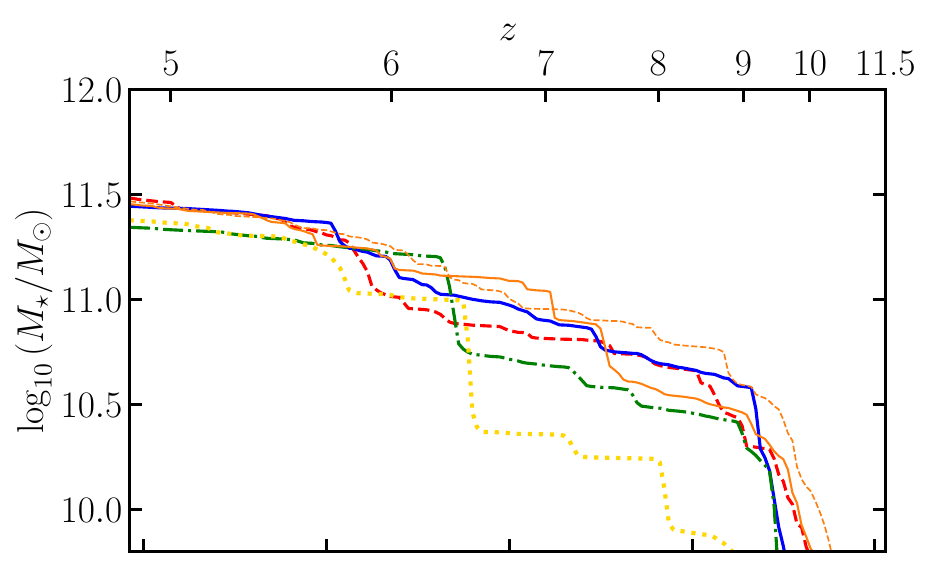}
        \hspace*{-0.3cm}
        \includegraphics[width=0.985\columnwidth]{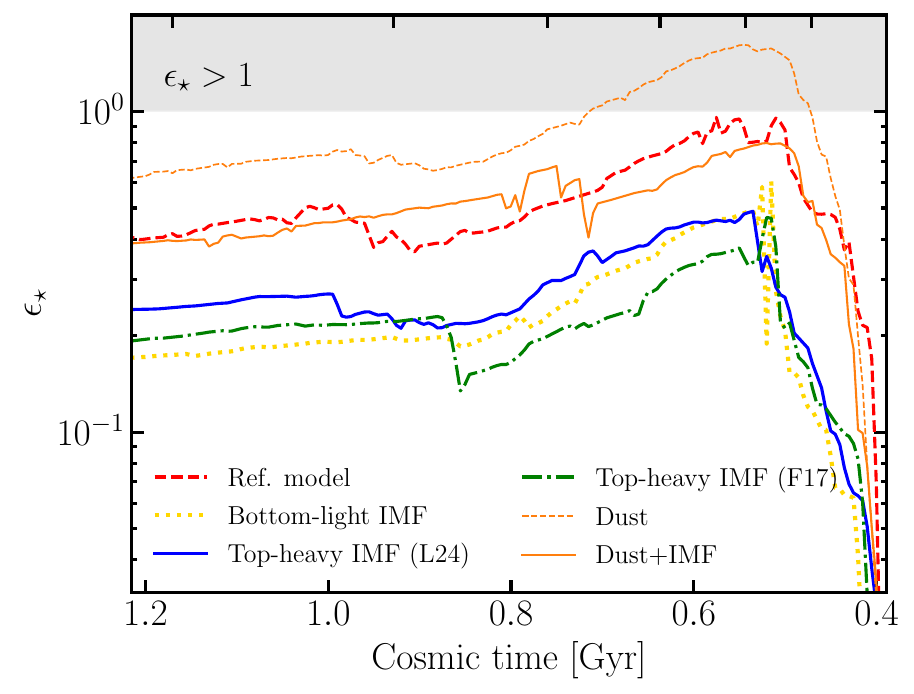}
        \caption{Upper panel: Stellar mass growth of galaxies in the same highest stellar mass bins as in Figure \ref{fg:sfh}, for our reference model IMF (red dashed line), the bottom-light (gold dotted line), and the \citet{lapi_2024} and \citetalias{fontanot_2017} top-heavy IMFs (blue solid and green dash-dotted lines). The orange dashed and solid lines show the cases with dust, and dust along with a top-heavy \citet{lapi_2024} IMF. Lower panel: Same as the upper panel, but for the star formation efficiency.}
        \label{fg:SFH_SFE_varIMF}
    \end{figure}

    Figure \ref{fg:IMFs} reports the various IMFs that we investigate in this Section. In particular, in addition to the canonical \citet{chabrier_2003} IMFs (see also \citealt{salpeter_1955, kroupa_2001}), we consider more exotic ones such as bottom-light and top-heavy, which have been constructed by tweaking the characteristic mass of the \citet{larson_1998} formula to $m_{\rm \star,c} = 0.1 \, M_\odot$ for the former, and the slope at high star masses to $\xi = -1.5$ for the latter. From the IGIMF library considered in \citetalias{fontanot_2017}, we focus on the IMF corresponding to a $\log_{10} \, {\rm SFR} = 3$, as the corresponding top-heavy shape provides the largest deviations from the canonical Chabrier IMF. We also note that, in principle, the $k_{\rm UV}$ conversion factor should vary with stellar age and metallicity, but we find that it has a weak dependence on these parameters in the ranges of interest to this work, stellar ages of $10 - 200$ Myr and metallicities $0.001-1 \, Z_\odot$ (e.g. \citealt{carnall_2023b, curtis_lake_2023, labbe_2023, tacchella_2023, topping_2024}).

    For the above-mentioned IMFs and input assumptions, we retrieve the following values of the UV luminosity-to-SFR conversion factors:
    \begin{equation}
        k_{\rm UV,Kroupa} = 2.1\times 10^{-10} \, M_\odot \, {\rm yr}^{-1} \, L_\odot^{-1} \, L_{\rm UV} \; ,
    \end{equation}
    \begin{equation}
        k_{\rm UV,bottom-light}^{\rm L24} = 1.1\times 10^{-10} \, M_\odot \, {\rm yr}^{-1} \, L_\odot^{-1} \, L_{\rm UV} \; ,
    \end{equation}
    \begin{equation}
        k_{\rm UV,top-heavy}^{\rm L24} = 1.7\times 10^{-10} \, M_\odot \, {\rm yr}^{-1} \, L_\odot^{-1} \, L_{\rm UV} \; ,
    \end{equation}
    \begin{equation}
        k_{\rm UV,top-heavy}^{\rm F17} = 1 \times 10^{-10} \, M_\odot \, {\rm yr}^{-1} \, L_\odot^{-1} \, L_{\rm UV} \; .
    \end{equation}
    The abundance matching procedure is then repeated by using the new conversion factors which lead to different SFR functions (see also Appendix \ref{app:SFRF_varIMF}).

    Interestingly, when assuming an input bottom-light IMF, the output SFRs would be a factor of $\sim 2$ lower than that predicted by the fiducial \citet{chabrier_2003}. Conversely, when adopting the top-heavy IMF from the \citet{lapi_2024} formalism, the SFRs would be lowered by the same factor, and by up to $\sim 50\%$ when adopting the top-heavy IMF from the \citetalias{fontanot_2017} formalism. Figure \ref{fg:SFH_SFE_varIMF} shows the stellar mass assembly (upper panel) and SFE (lower panel) in the most massive bin as in Figure \ref{fg:sfh}, for the variable bottom-light and top-heavy IMFs. Both a bottom-light and a top-heavy IMF could alleviate the need for extreme values of the SFEs and better accommodate the observed massive galaxies within $\Lambda$CDM, though still requiring peak values of the SFEs of around $30-40\%$. However, if we reinsert dust corrections in our non-canonical IMF models, we would still recover very high SFEs of $\epsilon_\star \gtrsim 0.7$ at $z\gtrsim 7$ (orange solid lines in Figure \ref{fg:SFH_SFE_varIMF}). Instead, as mentioned above, the reference model with only dust correction would yield unphysical SFEs greater than unity at $z\gtrsim 7$ (orange dashed lines in Figure \ref{fg:SFH_SFE_varIMF}). Thus, a non-canonical IMF appears to be a necessary ingredient to contain the SFEs of high redshifts, in particular in massive and dusty galaxies.

\section{Discussion}\label{sec:discuss}

In this work, we laid out a self-consistent and minimal data-driven approach to assign empirical SFRs to haloes and build the implied galaxy star formation histories. Instead of using observed UV LFs as outputs to test the model, we used them as input to infer SFRs mapped onto host dark matter haloes via abundance matching techniques. With the least possible assumptions and theoretical inputs, our methodology transparently allows us 1) to infer the efficiencies of star formation in haloes as emerging directly from the observed galaxy luminosities, and 2) to test the consequences of adding dust obscuration and/or IMF variations on empirical SFRs. In general, we find that high SFEs appear to be unavoidable within the framework of a $\Lambda$CDM Universe, independently of any stochasticity in the star formation histories or dust obscuration, although the inclusion of a variable IMF could reduce the SFEs by a factor of $\sim2$. Below we further discuss some implications of our results and how they compare to other works.

We showed in Figures \ref{fg:sfh} and \ref{fg:SFR_evo_data} that our method naturally predicts bursts of star formation characterised by extremely high SFEs following the sustained halo accretion histories characterising host haloes at early epochs. This finding aligns with what has been recently found in cosmological simulations by \citet{dekel_2025}, who reported feedback-free starbursts with massive galaxies at cosmic dawn. Similar results are also reported by the work of \citet{ferrara_2025b}, who found feedback-free SFRs in the pre-supernova phase with SFE $\epsilon_\star >0.4$. We also found that these galaxies are characterised by a number density of $\sim 1.8 \times 10^{-5} \, {\rm Mpc}^{-3}$ at redshift $z\sim 5-6$ for $M_\star \gtrsim 10^{11}\, M_\odot$, which lies far below the hard upper limit imposed by the dark matter halo mass function of $\sim 10^{-3} \, {\rm Mpc}^{-3}$ at the same redshifts (as shown in \citealt{boylan_kolchin_2023}), showing no tension between these objects and $\Lambda$CDM.

Another important caveat of our modelling is the lack of IR restframe data at redshift $z\gtrsim6$. We have shown that assuming an extinction law to account for the dust obscuration, may generate some overestimation of the measured SMF and further increase the already extreme SFEs above unity (dashed orange lines in Figure \ref{fg:SFH_SFE_varIMF}), although it could help in boosting the star formation histories of the very massive galaxies. We speculate that dust correction is a relevant factor in massive galaxies at the late phases of cosmic dawn, but less impactful at higher redshift and lower stellar masses. This hypothesis is also supported by the findings of several other works, that support a dust-free star formation scenario above $z\gtrsim10$ with a gradual transition period at redshift $z\sim 8-9$ (e.g. \citealt{ferrara_2022, ferrara_2023, ferrara_2025, algera_2023, cullen_2024}).

Finally, we studied how the galaxy SFE, for a given observed luminosity, can vary by assuming different IMFs. In particular, we found that a more exotic IMF, e.g. characterised by more massive stars, is expected to improve the above tension between the JWST high-luminosity objects and $\Lambda$CDM. This is due to the fact that more massive stars generally emit a higher amount of light per unit mass, producing higher luminosities at fixed galactic stellar masses. Indeed, as a result of this, we found that a top-heavy or bottom-light IMF yields SFEs a factor of $\sim 10-50\%$ lower with respect to the extreme values of SFEs $\epsilon_\star \sim 1$ suggested by the fiducial model, in good agreement with what reported in \citet{ziegler_2025} and \citet{fontanot_2026}. We found that our reference model with dust and/or the inclusion of a non-canonical IMF, still yields a clustering strength at large scales within $\pm 10-20\%$. Therefore, at the level of precision of the current data, galaxy clustering may not yet be able to efficiently break degeneracies among different models, but it can still be used as a consistency check for the validation of the mean correlation between stellar mass and host halo mass.

All in all, our findings suggest that the observed UV LFs necessarily translate into high SFEs when mapped against the halo accretion histories in a $\Lambda$CDM early Universe at $z\gtrsim 7$, in line with a variety of previous studies (e.g. \citealt{dekel_2023, ferrara_2023, lapi_2024, menci_2024, somerville_2025, prada_2026}). In turn, such high SFEs, when integrated in time, can generate massive galaxies close in stellar mass to those observed by JWST. Even in the presence of a non-canonical stellar IMF, the SFEs remain considerably high and close to unity if dust corrections are included. Beyond-$\Lambda$CDM cosmological models, making use of, for example, dynamical dark energy, could release some of the tension, although they would still require high SFEs (e.g. \citealt{menci_2026, comini_2026}).

\section{Conclusions}\label{sec:conclusions}

In this work, we have presented a flexible yet powerful data-driven method to estimate the star formation histories and efficiencies of the high-redshift galaxies detected by JWST. Instead of fitting the observed UV LFs through parametric SFEs, we have used these data as input in our framework to extract the implied SFRs, which we have mapped onto host dark matter haloes via abundance matching techniques with the halo accretion rates. Our SFRs are thus by design directly rooted in observations and as such can be considered an effective balance among any cooling, feedback and accretion, without requiring explicit modelling of any of these processes. By assigning SFRs to galaxies via our abundance matching relations along the main progenitors, we have built star formation histories, star formation efficiencies, and integrated stellar masses. In our transparent and minimalistic method based on abundance matching, SFEs simply emerge from the data rather than being a tuned input of the model, thus serving as a natural baseline to test burstiness, dust attenuation, or IMF variations.

Our main results can be summarised as follows:
\begin{itemize}
    \item The assumption of a monotonic relation between the galaxy SFR and dark matter halo accretion rate can robustly predict and match the bulk of the observed galaxy stellar mass function, supporting the view that the vast majority of galaxies above $z \gtrsim 5$ are mostly star-forming (Figures \ref{fg:cosmic_sfh} and \ref{fg:phi_mstar}).
    \item The star formation efficiency implied by our data-driven formalism is contained within $30\%$ up to $z\sim6$ but rapidly increasing to $\epsilon_\star \gtrsim 0.8$ at $z>9$ for $M_\star \gtrsim 5\times 10^{10}\, M_\odot$ (Figure \ref{fg:smhm_sfe}).
    \item We find that massive galaxies formed the majority of their stellar mass via a bursty star formation roughly 400 Myr after the Big Bang with extreme values of baryon conversion efficiencies (Figures \ref{fg:sfh} and \ref{fg:SFR_evo_data}).
    \item Our predicted star formation histories along the main progenitors generate a main sequence consistent with current measurements at these high redshifts (Figure \ref{fg:sfms}).
    \item Dust obscuration may play an important role in massive galaxies and can increase the total SFR by a factor of up to $\sim 2$ on average at redshift $5 \lesssim z \lesssim8$, with respect to the reference case without dust, implying SFEs greater than unity at $z\gtrsim 7$.
    \item Non-canonical IMF scenarios, like a top-heavy one (Figure \ref{fg:IMFs}), produce lower SFRs and SFEs at fixed luminosity, alleviating the pressing need for an extreme baryon conversion efficiency at $z>9$ (Figure \ref{fg:SFH_SFE_varIMF}).
\end{itemize}

In conclusion, all our results point to high SFEs, close to unity at $z\gtrsim 7$, even when accounting for non-standard IMFs, especially when correcting for dust extinction. Semi-empirical models represent a powerful data-driven approach for addressing key issues of galaxy evolution at high redshifts. Semi-empirical models can accurately predict several galactic properties, such as the star formation histories, galaxy abundances, and stellar mass-halo mass relation in line with observations, with only a minimal number of observed parameters as input. With the upcoming high-quality data from observatories, such as Euclid (e.g. \citealt{euclid_collaboration_2025_mission_overview}) and Rubin-LSST (e.g. \citealt{ivezic_2019}), the role played by data-driven models like ours will become even more relevant to extract key constraints on galaxy formation and evolution quantities in a self-consistent and flexible way (e.g. \citealt{lapi_2026}).

\begin{acknowledgements}
We warmly thank Y. Harikane for useful discussion on the UV luminosity function.We cordially thank R. Somerville, K. Glazebrook and L. Graziani for insightful discussions on the evolution of high-redshift galaxies. HF acknowledges support at Fudan University from the Shanghai Super Post-doctoral Excellence Program grant No. 2024008 and from the National Science Foundation of China (NSFC) grant No. W2533002. FS acknowledges partial support from the European Union’s Horizon 2020 research and innovation programme under the Marie Sk\l odowska-Curie grant agreement No. 860744 for the BiD4BESt project. AL acknowledges support from the Italian Research Center on High Performance Computing Big Data and Quantum Computing (ICSC), project funded by European Union - NextGenerationEU - and National Recovery and Resilience Plan (NRRP) - Mission 4 Component 2 within the activities of Spoke 3 (Astrophysics and Cosmos Observations). FY is supported in part by NSFC (grant Nos. 12133008, 12192220, and 12192223). MA is supported at the Argelander Institute f\"ur Astronomie through the Argelander Fellowship. LB acknowledges financial support from the German Excellence Strategy via the Heidelberg Cluster of Excellence (EXC 2181 - 390900948) STRUCTURES.
\end{acknowledgements}

\bibliographystyle{aa}
\bibliography{main}


\begin{appendix}

\section{High-redshift star formation rate function}\label{app:highz_SFRF}

\begin{figure}
    \centering
    \includegraphics[width=\columnwidth]{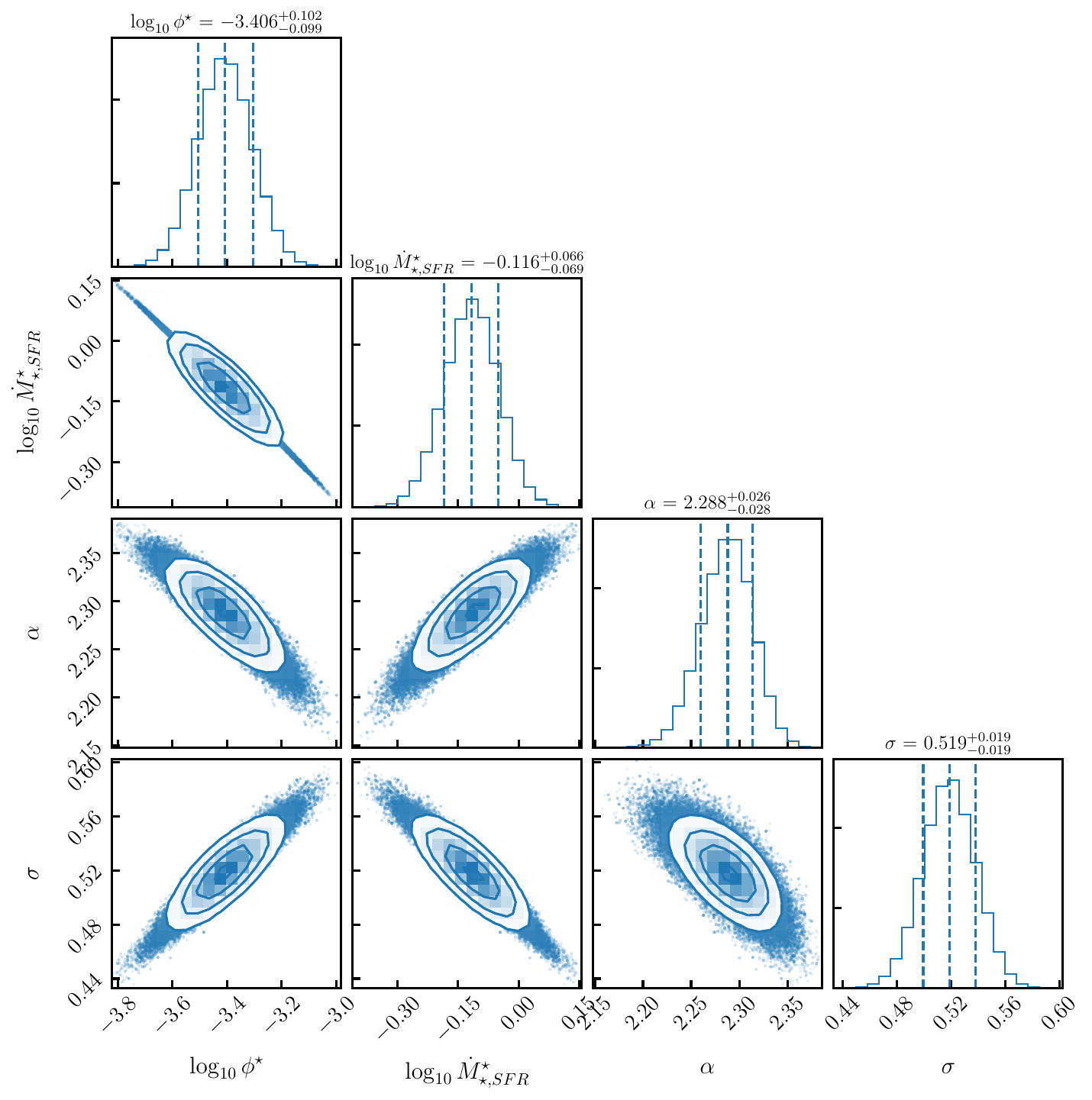}
    \caption{Poster distribution of the fitting parameters to Equation (\ref{eq:saunders}) for the star formation rate function at redshift $z=10$.}
    \label{fg:corner_z10}
\end{figure}

We fit the collection of data for the SFR function, as described in Section \ref{sec:res_sfr_har} (Figure \ref{fg:phi_sfr_data}), with a modified Schechter function following \citet{saunders_1990}
\begin{equation}\label{eq:saunders}
\begin{split}
    \phi(\dot{M}_{\rm \star,SFR}) \mathrm{d} & \log \dot{M}_{\rm \star,SFR} = \dot{M}^\star_{\rm \star,SFR} \bigg( \frac{\dot{M}_{\rm \star,SFR}}{\dot{M}^\star_{\rm \star,SFR}} \bigg)^{1-\alpha} \\
    \exp & \bigg[ - \frac{1}{2\sigma^2} \log^2_{10} \bigg( 1 + \frac{\dot{M}_{\rm \star,SFR}}{\dot{M}^\star_{\rm \star,SFR}} \bigg) \bigg] \mathrm{d} \log \dot{M}_{\rm \star,SFR} \;\;\; .
\end{split}
\end{equation}
The best-fitting parameters of the SFR function to Equation (\ref{eq:saunders}) at each redshift are shown in Table \ref{tab:SFR_func_fit_params}. Figure \ref{fg:corner_z10} shows, as an example of robustness of our fit, the posterior distributions of the parameters fitted at redshift $z=10$, and we obtained also similar corner plots for the fits at any other redshifts considered in this work.  We also fit the redshift evolution of each parameter of the Saunders formula, using minimum likelihood method, finding the following analytic redshift-dependent fits
\begin{equation}
    \begin{split}
        \log_{10} \phi^\star (z) & = -3.43 + 1.59 \cdot \log_{10}(1+\Tilde{z}) \\
        & -6.39 \cdot \log^2_{10}(1+\Tilde{z}) + 4.83 \cdot \log^3_{10}(1+\Tilde{z}) \; ,
    \end{split}
\end{equation}

\begin{equation}
    \begin{split}
        \log_{10} \dot{M}^\star_{\rm \star,SFR} (z) & = 0.71 + 2.62 \cdot \log_{10}(1+\Tilde{z}) \\
        - 11.06 & \cdot \log^2_{10}(1+\Tilde{z}) + 8.69 \cdot \log^3_{10}(1+\Tilde{z}) \; ,
    \end{split}
\end{equation}

\begin{equation}
    \begin{split}
        \alpha (z) = 2 & - 2.48 \cdot \log_{10}(1+\Tilde{z}) \\
        & + 10.8 \cdot \log^2_{10}(1+\Tilde{z}) - 9.23 \cdot \log^3_{10}(1+\Tilde{z}) \; ,
    \end{split}
\end{equation}

\begin{equation}
    \begin{split}
        \sigma (z) = 0.22 & - 1.52 \cdot \log_{10}(1+\Tilde{z}) \\
        & + 5.88 \cdot \log^2_{10}(1+\Tilde{z}) - 4.59 \cdot \log^3_{10}(1+\Tilde{z}) \; ,
    \end{split}
\end{equation}

where $\Tilde{z} = z-z_0$ and $z_0=6$.

\begin{table}
    \caption{Best-fitting parameters to Equation (\ref{eq:saunders}) for the SFR function at redshifts $z\geq 6$.}
    \centering
    \begin{tabular}{ccccc}
        \hline
        $z$ & $\log_{10} \phi^\star$ & $\log_{10} \dot{M}^\star_{\rm \star,SFR}$ & $\alpha$ & $\sigma$ \\
        \hline
        $6$ & $-3.425^{+0.027}_{-0.027}$ & $0.712^{+0.027}_{-0.027}$ & $1.995^{+0.002}_{-0.002}$ & $0.220^{+0.012}_{-0.012}$ \\
        \hline
        $7$ & $-3.425^{0.007}_{0.007}$ &$0.712^{0.004}_{0.004}$ & $1.995^{0.003}_{0.003}$ & $0.19^{0.001}_{0.001}$ \\
        \hline
        $8$ & $-3.414^{0.051}_{0.052}$ &$0.367^{0.038}_{0.037}$ & $2.191^{0.007}_{0.007}$ & $0.305^{0.014}_{0.014}$ \\
        \hline
        $9$ & $-4.148^{0.104}_{0.111}$ &$0.336^{0.058}_{0.055}$ & $2.525^{0.054}_{0.054}$ & $0.412$ \\
        \hline
        $10$ & $-3.406^{0.102}_{0.099}$ & $-0.116^{0.066}_{0.069}$ & $2.288^{0.026}_{0.028}$ & $0.519^{0.019}_{0.019}$ \\
        \hline
        $11$ & $-3.917^{0.125}_{0.128}$ &$0.229^{0.082}_{0.083}$ & $2.278^{0.036}_{0.039}$ & $0.399^{0.023}_{0.024}$ \\
        \hline
    \end{tabular}
    \label{tab:SFR_func_fit_params}
\end{table}

\section{Star formation rate function for different initial mass functions}\label{app:SFRF_varIMF}

\begin{figure*}
    \centering
    \includegraphics[width=0.95\textwidth]{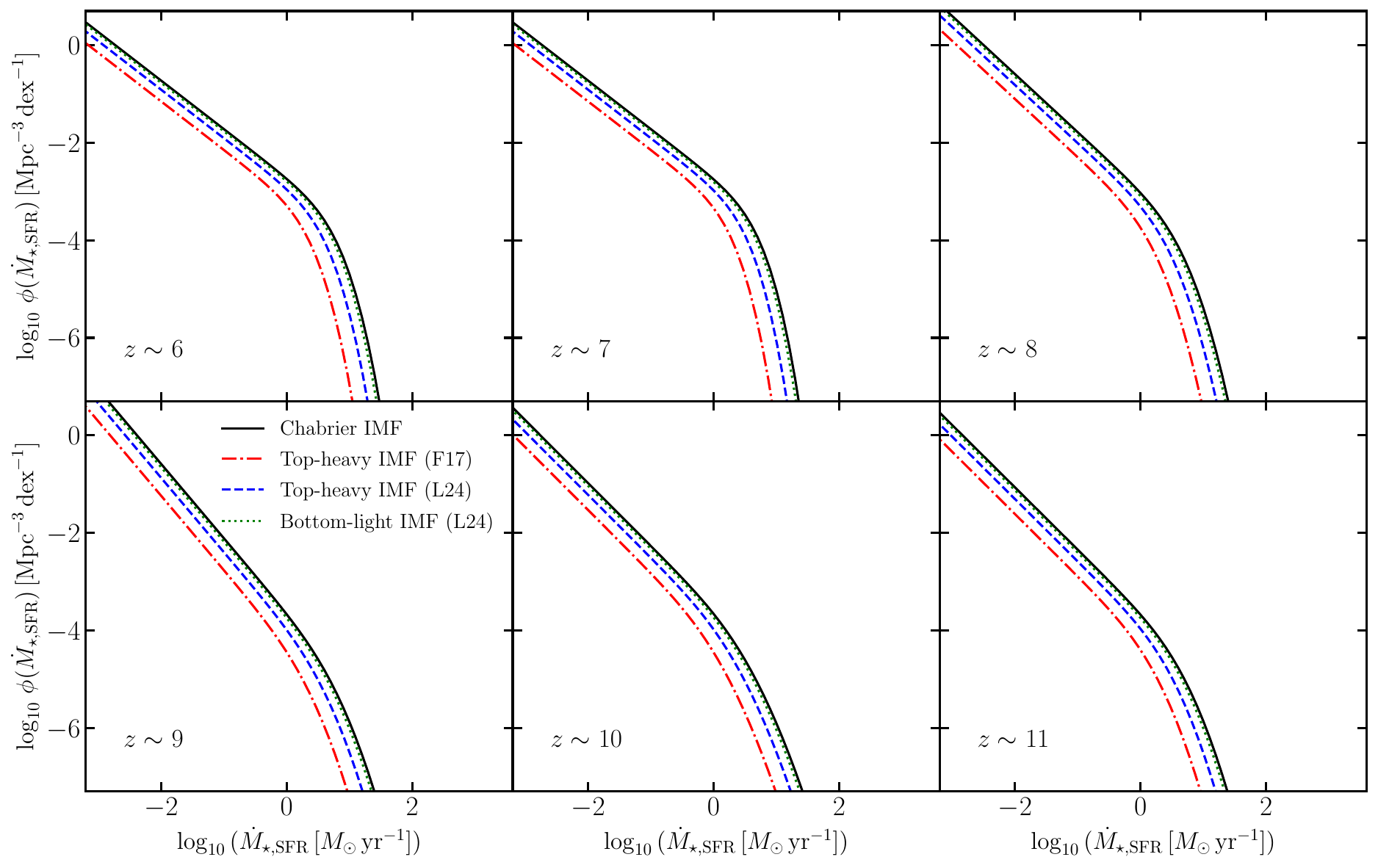}
    \caption{Star formation rate function from redshifts $z = 6$ to $11$ for different assumed initial mass functions: canonical (black solid lines; \citealt{chabrier_2003}), top-heavy F17 (red dash-dotted lines; \citetalias{fontanot_2017}), top-heavy L24 (blue dashed lines; \citealt{lapi_2024}), and bottom-light (green dotted lines; \citealt{lapi_2024}). The red dash-dotted lines refer to the \citetalias{fontanot_2017} IGIMF for ${\rm SFR} = 10^3 \, M_\odot/{\rm yr}$.}
    \label{fg:phi_sfr_varIMF}
\end{figure*}

Figure \ref{fg:phi_sfr_varIMF} shows the SFR function for the different IMFs we explore in Section \ref{sec:res_imf}. For each input IMF, we repeat the abundance matching by building a new SFR function depending on the $L_{\rm UV}$-to-SFR conversion corresponding to the assumed IMF.

\section{Dust attenuation laws}\label{app:dust_laws}

We first employ the \citet{charlot_fall_2000} attenuation law, according to which the intrinsic luminosity is correlated to the observed luminosity with the following power law
\begin{equation}\label{eq:charlot_fall}
    \begin{split}
        L_{\rm int} &= L_{\rm obs} \times e^{\tau_{\lambda}} \; ,\\
        \uptau_\lambda &= \tau_\lambda^{\rm ISM} \cdot \bigg( \frac{\lambda}{5500 \, \AA} \bigg)^{-n} + \tau_{\lambda}^{\rm BC} \cdot \bigg( \frac{\lambda}{5500 \, \AA} \bigg)^{-n} \, .
    \end{split}
\end{equation}
where $\lambda$ is the observation wavelength and $(\tau_\lambda^{\rm BC}, \tau_\lambda^{\rm ISM}, n) = (1,0.5,0.7)$.

We secondly explore the \citet{calzetti_2000} attenuation law
\begin{equation}\label{eq:calzetti}
    L_{\rm int} = L_{\rm obs} \times 10^{0.4 A_{\lambda}} \, ,
\end{equation}
with $A_{\lambda} = k(\lambda) \cdot E(B-V)$. Here, $E(B-V)$ is the colour excess and $k(\lambda) = 2.659 (-1.857 + 1.040 / \lambda) + R_V$, with $R_V = 4.05$.

Lastly, we also include the modified Calzetti law as presented in \citet{salim_2018}
\begin{equation}\label{eq:calzetti_mod}
\begin{split}
    k(\lambda) & = k_{\rm \lambda, Cal} \frac{R_{V,{\rm mod}}}{R_{V,{\rm Cal}}} \bigg( \frac{\lambda}{5500 \AA} \bigg)^\delta + D_\lambda \, , \\
    R_{V,{\rm mod}} & = \frac{R_{V,{\rm Cal}}}{ (R_{V,{\rm Cal}} + 1) (4400/5500)^\delta - R_{V,{\rm Cal}} } \, , \\
    \delta & = -0.45 + 0.19 (\log_{10}(M_\star/M_\odot) -10) \, ,
\end{split}
\end{equation}
where $R_{V,{\rm Cal}} = 4.05$ and $D_\lambda$ is the Drude profile UV bump (\citealt{fitzpatrick_massa_1986}).

\begin{figure}
    \centering
    \includegraphics[width=\columnwidth]{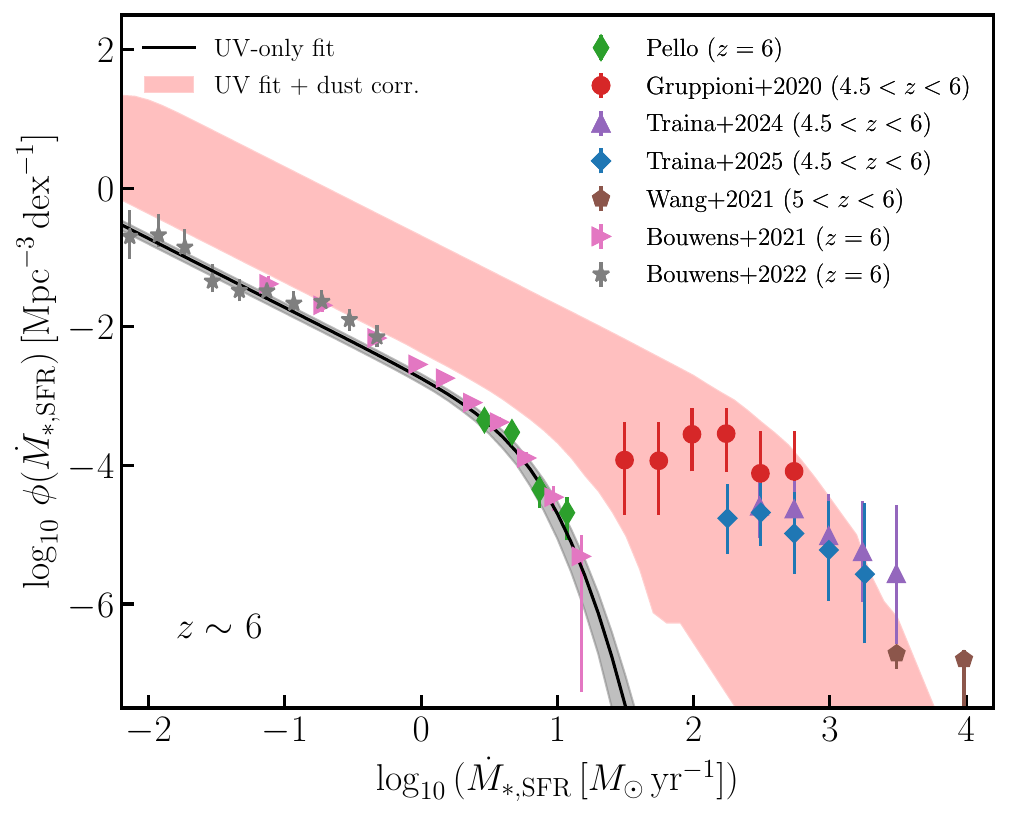}
    \caption{Star formation rate function at redshift $z = 6$ from the UV-only fit and by including the dust correction, compared to several observations both in the UV and IR bands.}
    \label{fg:phi_sfr_test_IR}
\end{figure}

We note that the above correction for dust depends on the specific observation wavelength of the sources. We checked that by applying the correction from the above recipes within the wavelengths of interest the bright end of the galaxy LF matches the observed one in the IR band, but naturally overestimates the faint end where the dust content should be significantly lower (Figure \ref{fg:phi_sfr_test_IR}).

\end{appendix}

\end{document}